\documentclass[aas2pp4]{aastex}

\begin{document}

\def\arcdeg{\hbox{$^\circ$}}
\def\arcmin{\hbox{$^\prime$}}
\def\arcsec{\hbox{$^{\prime\prime}$}}
\def\ltsim{\mathrel{\hbox{\rlap{\hbox{\lower4pt\hbox{$\sim$}}}\hbox{$<$}}}}
\def\gtsim{\mathrel{\hbox{\rlap{\hbox{\lower4pt\hbox{$\sim$}}}\hbox{$>$}}}}
\def\h{\hbox{$^h$ \ }}
\def\pp{\hbox{$P^\prime$}}
\def\micron{\hbox{$\mu$m}}

\title{The Canada-UK Deep Submillimeter Survey VII: Optical and Near-Infrared Identifications for the 14\h Field\altaffilmark{1}}

\author{T.M.A. Webb\altaffilmark{2,3,4}, S.J. Lilly\altaffilmark{5}, D.L. 
Clements\altaffilmark{6}, S. Eales\altaffilmark{6},  
M. Yun\altaffilmark{7},  M. Brodwin\altaffilmark{2,4}, L. Dunne\altaffilmark{6}, W.K. Gear\altaffilmark{6}}

\altaffiltext{1} {Based on observations made with the NASA/ESA Hubble Space Telescope, obtained from the Data Archive at the Space Telescope Science Institute, which is operated by the Association of Universities for Research in Astronomy, Inc., under NASA contract NAS 5-26555. These observations are associated with proposal numbers \# 5090, \#5109, \#5449, \#8162.} 
\altaffiltext{2}{Department of Astronomy and Astrophysics, University of Toronto, 60 St George St, 
Toronto, Ontario, M5S 1A1, Canada}
\altaffiltext{3} {Sterrewacht Leiden, Neils Bohrweg 2, Leiden 233CA, The Netherlands} 
\altaffiltext{6}{Department of Physics and Astronomy, Cardiff University, P.O. Box 913, Cardiff, CF2 
3YB, UK} 
\altaffiltext{5}{Institut f\"{u}r Astronomie, ETH H\"{o}nggerberg, HPF G4.1, CH-8093, Z\"{u}rich,
Switzerland}
\altaffiltext{7}{Department of Astronomy, University of Massachusetts, Amherst,
MA 01003, USA}
\altaffiltext{4}{Visiting Astronomer, Canada-France-Hawaii Telescope, Operated by the National 
Research Council of Canada, the Centre de la Recherche Scientifique de France, and the University of 
Hawaii.}

\begin{abstract}

We present the multi-wavelength identifications for 23 sources in the Canada-UK Deep Submillimeter 
Survey (CUDSS) 14\h field.  The identifications have been selected  on the basis of radio and near-infrared 
data and we argue that, to our observational limits,  both are effective at 
selecting the correct counterparts of the 
SCUBA sources.   We discuss the properties of these identifications and find that they are very red in 
near-infrared color, with many classified as Extremely 
Red Objects, and  show disturbed morphologies.  Using the entire CUDSS catalogue of 50 sources  we use a combination of spectroscopic redshifts (4 objects), 1.4GHz-to-850$\mu$m flux ratio redshift estimates (10 objects), and redshift lower-limits based on non-detections at 1.4GHz (the rest of the sample) to estimate a lower-limit on the median redshift of 
the population of $z_{med} >$ 1.4.   Working from simple models and using the properties 
of the secure identifications, we discuss general and tentative 
constraints on the redshift distribution and the expected colors and magnitudes of the entire population.  

\end{abstract}


\section{Introduction}

The submillimeter surveys of the last five years, using the Submillimeter Common-User Bolometer Array (SCUBA) and the Max-Planck Millimeter Bolometer array (MAMBO) \citep{sma97,bar98,hug98,eal99,bor02,cow02,sco02,dan02}  have revealed a population of high-redshift objects which  play an important role in the formation and evolution of galaxies.
Though relatively rare ($\sim$0.5 arcmin$^{-2}$ at $S_{850{\mu}m}>$ 3 mJy) these systems have extreme 
individual luminosities ($>$ 10$^{12}$ L$_{\odot}$)  
and are responsible for  $\sim$ 20\% of the far-infrared background at 850\micron \ (to $S_{850{\mu}m} >$ 3 mJy) \citep{bla99,bar99a,cow02,sco02,web03b}.   This is in striking contrast to the relative  unimportance of 
similar far-infrared bright galaxies in the local universe (see \citet{san96} for an extensive review),
and indicates substantial evolution in this population with redshift.  The high star formation rates of these galaxies of
$\sim$100-1000 M$_{\odot}$/year, are  sufficient  to produce a massive galaxy over a dynamical timescale, and make them strong candidates for the progenitors of   local elliptical galaxies.
   However, 
there are still many unanswered questions 
regarding the nature of these objects, their redshift distribution,  and their relationship to other  high-redshift populations.

A key issue, but one of the hardest to address, is the nature of the obscurred energy source. Determining the fraction of energy produced by  star-formation and active galactic nuclei (AGN) is crucial for understanding how these systems fit into theories of galaxy evolution (e.g. \citet{arc02}). Though a number SCUBA sources which have been studied in detail show signs of AGN activity, and indeed, all SCUBA sources may be home to obscured AGN,  the current evidence points toward star-formation dominated systems \citep{fra98,fra99,bar99b,bar01,sma02,ivi02,was03,alm03,ale03}. Even once the relative contribution of an AGN to the submillimeter flux is known,   inferring a star formation rate is still difficult.  The primary uncertainty results from the poorly understood dust temperature in these systems and though we may attempt to calibrate this using $z\sim$0  ULIRGs \citep{dun00a,yun02}, there is growing evidence that these local objects may not be representative of the high-redshift SCUBA sources (e.g. \citet{ivi01,lut01,fra03}). An understanding of the energy production and radiation mechanisms in these systems will require extensive follow-up observations, with a level of detail similar to those of \citet{lut01},\citet{led02},\citet{gen03},and \citet{fra03} for many systems. The speed with which we can carry-out these observations is limited by current technology but this will change dramatically with future facilities such as ALMA, JWST, and the LMT.  

Though these systems certainly have sufficient star formation rates to rapidly form an elliptical galaxy, it is less clear that their spatial density (e.g. \citet{fox02})  and  clustering properties \citep{ivi00a,sco02,web03b} are consistent with local massive elliptical galaxies. This would be an evolutionary link between the two populations, but requires a complete determination of the redshift distribution, and in particular, for a  measure of the clustering strength, significant sky coverage.   
Such measurements would also shed light on the relationship between these objects and other high-redshift populations.  Though  the Lyman-break galaxies have very little direct overlap with the bright submillimeter population ($S_{850{\mu}} \gtsim$ 3 mJy), they are likely present at significant numbers in the fainter  $S_{850{\mu}m}$ counts \citep{pea00,cha00,web03a}.  There is also tentative evidence that the LBGs and SCUBA sources may trace the same large scale structure \citep{web03a}, perhaps forming a mass sequence in which SCUBA sources represent the most massive star forming systems at high-redshift\citep{mag01,gra01,gen03}, or alternatively, they may be  related through time, as systems evolve from a dusty, luminous stage of intense star formation though a relatively dust-free and perhaps extended phase of moderate star formation. \citep{sha01,ivi02}. 

The determination of the redshift distribution is hindered by the  uncertainties in the source positions, due relatively large beam size of  the JCMT at 850{\micron}.  Positions are generally secured  through  radio detections which biases  the  measured redshifts to $z\lesssim$ 3 \citep{cha03}. A further possible complication is the different flux limits reached by different surveys, ranging from the sub-mJy limits of the cluster surveys \citep{sma97,cha02} to the very bright ($S_{850{\micron}} >$ 8mJy) limits of the '8 mJy survey' \citep{sco02}. There are  indications that the brighter SCUBA sources may lie at predominantly  higher-redshifts than the fainter systems \citep{cha02,ivi02}. If so, relatively deep surveys such as this one would be biased to lower-redshift sources compared to shallower programs such as the ``8 mJy Survey'' \citep{sco02,fox02,ivi02}.

The resolution of all of these issues requires the careful and extensive follow-up of a wide variety of sources, over a range of flux levels and selection techniques. Unfortunately, the first step of identifying the optical counterparts of these systems is not trivial and  many  SCUBA sources catalogued in current surveys  have
no solid identification \citep{fox02,sma02,ivi02,web03b}.  This is a result of  the uncertainty in the submillimeter positions and  the faint and red nature of these objects at optical wavelengths (e.g. \citet{fra00,lut01,dunl02,dan02}). Hence, more than one candidate identification often  lies  within a given submillimeter error radius. Even once counterparts have been securely identified they are  typically so faint that optical and near-IR spectroscopic redshifts can only be measured with 8-m class telescopes (if at all) and  large investments of observing time are required (e.g. \citet{bar99b,cha03}.  This situation will improve with the commissioning of facilities which will measure redshifts through  masers, CO lines, and far-IR photometry, and will not require bright optical/NIR counterparts  \citep{com99,hug00,fra01,tow01,hug03,are03}

It is because of the empirical correlation  between  radio and far-infrared flux \citep{car99,dun00a,yun02}, and the excellent 
resolution of large radio arrays,  that radio  mapping currently offers 
the best chance to secure positions of 
the sources.  Moreover, the extremely low surface density of radio sources makes chance coincidences between the radio and submillimeter populations highly unlikely.    Unfortunately, at the current typical flux limits radio detections are 
biased to redshifts of z $\lesssim$ 2-3  
and, depending on the redshift distribution of the SCUBA sources, 
may  miss a substantial fraction of the population \citep{eal00,cha02,ivi02,web03b}. 
NIR imaging is crucial as  these  objects  may be  
expected to be very red due to the presence of dust (but also see \citet{tre99}).  Indeed, a significant fraction of SCUBA sources 
have been associated with  Extremely Red Objects (EROs) or Very Red Objects (VROs) \citep{dey99,gea00,sma99,fox02,ivi02,dan02,dunl02}. As in the radio, the relatively low  surface 
density of red objects, compared to optically selected populations, reduces the frequency of chance positional coincidences.

In this paper  we explore the multi-wavelength properties and redshift distribution of submillimeter sources 
detected in the Canada-UK Deep 
Submillimeter Survey (CUDSS). We present the identifications for sources in the 14\h field and discuss the properties of the entire catalogue. Here we briefly re-cap the previous papers which have resulted from this survey. The submillimeter  
data are discussed in detail in 
\citet{eal99,eal00} (Papers I and IV) and  \citet{web03b} (Paper VI).  The first  identifications were presented in \citet{lil99} (Paper II).  One 
of our brightest sources CUDSS 14.1, which 
has been observed extensively at other 
wavelengths, is discussed in \citet{gea00}(Paper III).  A discussion of the submillimeter properties of Lyman-break 
galaxies in the survey fields may be found in 
\citet{web03a} (Paper V). The optical/NIR identifications of the 3\h  field are presented in Clements et al 
(in preparation).

 This paper is organized as follows.
In \S 2 we outline the multi-wavelength data which we, and others, have obtained over this field area.  \S 
3 explains our procedure for identifying counterparts  and presents the identifications. In \S 4 we 
discuss the properties of individual sources. 
  In \S 5 contains a discussion of  our results in detail.  
Finally, in \S 6 we summarize our results.  We assume a flat, $\Lambda$=0.7 cosmology and H$_\circ$=72 km/s/Mpc throughout.

\section{The multi-wavelength data}

\subsection{The submillimeter observations}

The CUDSS Survey consists of two primary fields: the 14\h field (48 arcmin$^2$) and the 3\h field 
(60 arcmin$^2$).
  These fields are contained within the Canada-France Redshift Survey (CFRS) fields CFRS14+52 and 
CFRS03+00 respectively.    The 
submillimeter data were obtained using SCUBA on the JCMT over many observing runs from 1998 to 2001.  The details of the submillimeter observations  are 
discussed in Papers IV and VI. Paper VI also contains the complete 3\h field catalogue of 27 sources. 

In Paper IV we presented a source catalogue for the 14\h field of 19 objects with $S/N \geq$ 3.  Since 
then 
we have acquired new submillimeter data, primarily located on the upper strip of the field which, previously, had 
not been imaged to the same 
depth as the rest of the field.  Through the  addition of 
these data four new sources were detected, all at  $S/N \sim$ 3.    These new observations  marked the 
completion of the submillimeter survey of this field and 
we now present our final  catalogue in Table 1.  Note that sources 14.1 through 
14.19 are identical to the sources listed in Paper IV.

\subsection{Radio and ISO Data}

Most of the $14^h$ field has been mapped at 1.4 GHz and 5 GHz 
by Fomalont et al. (1991) to a depth of $\sim 16~\mu$Jy and 
2.5 $\mu$Jy, respectively.  Data at 1.4 GHz also exists for the 3\h field (Yun personal communication; Ivison personal communication) and was  discussed in Paper VI.
Taking advantage of the improved wide field imaging algorithms in the
AIPS software system, M. Yun re-calibrated the archival VLA data by
Fomalont et al. and obtained a new continuum image with about twice
the improved resolution ($\theta_{FWHM}\sim 4\arcsec$) using a
robust weighting of visibilities.  The final
dynamic range limited noise in the 14\h image is $14 \mu$Jy,
similar to that of Fomalont et al. As a comparison to other surveys we note that the ratio of the submillimeter to radio 1$\sigma$ noise level (per beam) in this survey is 1mJy/14$\mu$Jy $\sim$ 70, while for the HDF it is 0.4mJy/7.5$\mu$Jy $\sim$50 \citep{hug98,ric00} and for the '8 mJy survey' it is 2mJy/4.8$\mu$Jy $\sim$ 400 \citep{sco02,ivi02}. 

In addition to the improved angular resolution,  completely
different noise characteristics and a different treatment of
a serious imaging problem make the examination of the newly 
reduced 1.4 GHz image worthwhile.  Because these VLA observations 
are done in the continuum mode, both the 1.4 GHz and the
5 GHz images suffer from beam smearing (chromatic aberration) --
cross-correlation of signal with different wavelengths with the
bandpass causes a loss of amplitude and displacement of source 
position in a radial direction from the phase center (see Bridle
\& Schwab 1988).
This problem is particularly severe for the SCUBA sources because
they are located mostly outside the central 150{\arcsec} region
which is relatively problem-free.  Fomalont et al. dealt with
this problem by summing all flux over the radially elongated source 
regions (E. Fomalont, private communication in 2000).  Recovering
the total flux this way requires summing over more than one beam
areas (thus increasing noise) and may not fully account for the
amplitude de-correlation.  Instead, we have measured the peak
amplitudes from the image and then recovered the source flux
using the correction factor that depends on the radial distance
from the phase center using Eq.~13-24 in Bridle \& Schwab.
The new location and the recovered flux density for the SCUBA
sources in the 14$^h$ field are given in Table~\ref{radiotab}.
The new 1.4 GHz image does not yield any new detections of the
SCUBA sources within the search radius of 8{\arcsec}, but we can derive a more accurate upper limits
after properly taking into account the effects of beam smearing.

Both SCUBA fields have been imaged by the Infrared Space Observatory (ISO).  The 
data 
covering the 14\h field at 7\micron \ and 15\micron \ are discussed in detail 
in \citet{flo99a,flo99b}.  The data covering the 3\h field are not yet published and have been provided to us by 
Hector  Flores.

\subsection{Existing Optical Data: The CFRS and HST}

The CFRS fields have been extensively studied at optical wavelengths.  In addition to the original 
$UBVIK$ photometry and spectroscopic 
redshifts ($z\leq$ 1.3) of the CFRS itself  \citep{lil95b,ham95a},  Hubble Space Telescope (HST) $VI$ imaging was 
undertaken for  
CFRS morphology follow-up work  and further archival observations were available (see footnote 1, first page). 

\subsection{New HST Data}

 As part of  follow-up work for this survey we acquired three new $I$-band  HST 
pointings for each CUDSS field. These data were reduced and calibrated using the general HST reduction pipeline and  reach an $I_{AB}$ depth of $\sim$26.0 mags. These data, combined with the existing HST data described above, cover 19/23 objects in the 14\h field, and 19/27 objects in the 3\h field.

\subsection{New Near-Infrared Data: Kitt Peak and CFHTIR}

The $K$-band imaging of the CFRS was extremely shallow  and  covered only 1/3 the CFRS fields 
fields. Thus, we obtained 
new $K$-band imaging  using IRIM on the Mayall-4m at Kitt Peak  and 
 using the CFHTIR camera 
on the CFHT.  The CFHTIR data have better  pixel resolution than the IRIM data, with  0.207 arcsec/pixel, 
versus 0.6 arcsec/pixel, as well as 
improved sky quality.  The Kitt Peak data reach a depth of $K_{AB}\sim$21.5 mags and the CFHTIR data 
reach $K_{AB}\sim$22.5-23.0 mags, over most of the image.  Both data sets cover roughly 2/3 of the 14\h  and 
3\h fields. 

\section{The Multi-wavelength Identifications }

\subsection{The Identification Procedure}

We have chosen a  method which selects
counterparts  on the  basis of positional  coincidence (\citet{dow86}; Paper II) as follows.  The probability that an 
object, 
physically unrelated to a given SCUBA 
source,     will lie within a distance $r$ of said  source can be 
described by $P=1-exp(-{\pi}nr^2)$, where $n$ is the surface density of galaxies as bright or brighter 
than the candidate identification. Defined in this way, the lower the $P$ value, the higher the statistical significance of the identification.

As pointed out in Paper II, using the surface density of objects as bright or brighter than the candidate identification inherently underestimates the probability of random associations. This is because one    
effectively searches many independent galaxy samples simultaneously, and the chance that one of these 
samples will produce an identification of high 
significance is 
increased.  We must, therefore, calculate a new 
quantity $P^{\prime}=\alpha P$, where 
$\alpha$ is determined from Monte-Carlo simulations of our identification procedure. We use  randomly chosen 
positions in place of real SCUBA positions,  which yields an empirical estimate of the 
probability of finding a galaxy of a given magnitude and color within the search radius of a random 
position.  \

These $P^\prime$ values should be interpreted in the following way.   If we assume that SCUBA sources 
are completely unrelated to the 
optical/near-IR population within which we search for identifications, we would statistically expect to select 10 identifications 
with  $P^\prime \leq $ 0.1 for every 100 
SCUBA sources.  Thus, $P^\prime$ for a single identification should 
 be interpreted with respect to the 
entire sample, rather than in isolation. 

 We have empirically estimated our search radius by placing and recovering fake SCUBA sources in our 
submillimeter map, and through Monte-Carlo simulations of the data (Paper IV). We found  the 
positional offset between the measured position of an object and its true position was  
$\leq$ 8\arcsec \  90-95\% of the time, and  therefore we use this as our search radius.  This is a larger search 
radius than used by some  groups (e.g. 
\citet{bar99b,sma02}) though we note that the peak of our offset distribution lies at a smaller radius of 
2-4\arcsec \ (also see \citet{hog01}).

\subsection{The Radio and Mid-Infrared Identifications}

In Paper IV we presented five radio  identifications for the 14\h field and these are listed in Table \ref{radiotab}, along with their $P$ statistics.  At these levels of significance we do not expect any of these five identifications to be the result of random 
SCUBA-radio 
associations. In Table \ref{radiotab} we also include a possible radio counterpart for source 14.19 which lies beyond our search radius. As we expect 1-2 identifications (for our sample size) with offsets from the 
SCUBA position of  $>$ 8\arcsec,  this identification should be considered but is by no means secure. 

The ISO identifications were also discussed in Paper IV  and are summarized in Table \ref{isotab}.  
Three sources have been detected in the 
mid-IR, two of which (14.13, 14.18) are also coincident with 
radio emission. The third detection (14.17)  has 
a very large offset (10.3\arcsec) and is therefore not  a secure  identification.  \

\subsection{The Near-Infrared Selected Identifications}

In Paper II the identifications were chosen  from an $I$-selected population.  In this work we have improved our algorithm by including $K$ information as follows.  
For a given $K$-selected galaxy within the positional error radius of a SCUBA source we compute the 
surface density of galaxies as bright or brighter than its 
$K$-band  magnitude and as red or redder than its $(I-K)$ color, directly from our  CFHTIR mosaic image and the 
CFRS optical data.  Because we estimate this directly from the follow-up data field-to-field 
variations in density are taken into account. \

Using the $(I-K)$ 
color is advantageous for two reasons: (1) Since  red galaxies have a lower surface density than galaxies 
of more 
moderate color the chance of random coincidences is decreased, and (2) Because SCUBA 
sources are by their nature very dusty we may begin with the  reasonable assumption that they have redder 
colors than the typical galaxy.  This second 
point is somewhat contentious since bright submillimeter flux does not necessarily guarantee red optical/NIR 
colors. In fact,  \citet{tre99} has shown 
that some low-redshift ULIRGs have surprisingly blue optical colors when placed  at high-redshift.   \

$P^\prime$ was calculated  for each $K$-selected galaxy within 8\arcsec \  of each SCUBA source. This was done blindly with respect to the radio and mid-infrared data.  All 23 SCUBA sources had at least one candidate identification within their submillimeter  error radius (i.e. 
there were no demonstratable ``empty fields''), and  there 
were typically 1-3 possibilities. Table \ref{idtab} lists the best identification (i.e. the one with the lowest $P^\prime$ value) for each source.

\begin{figure} 
\epsscale{1.0}
\plotone{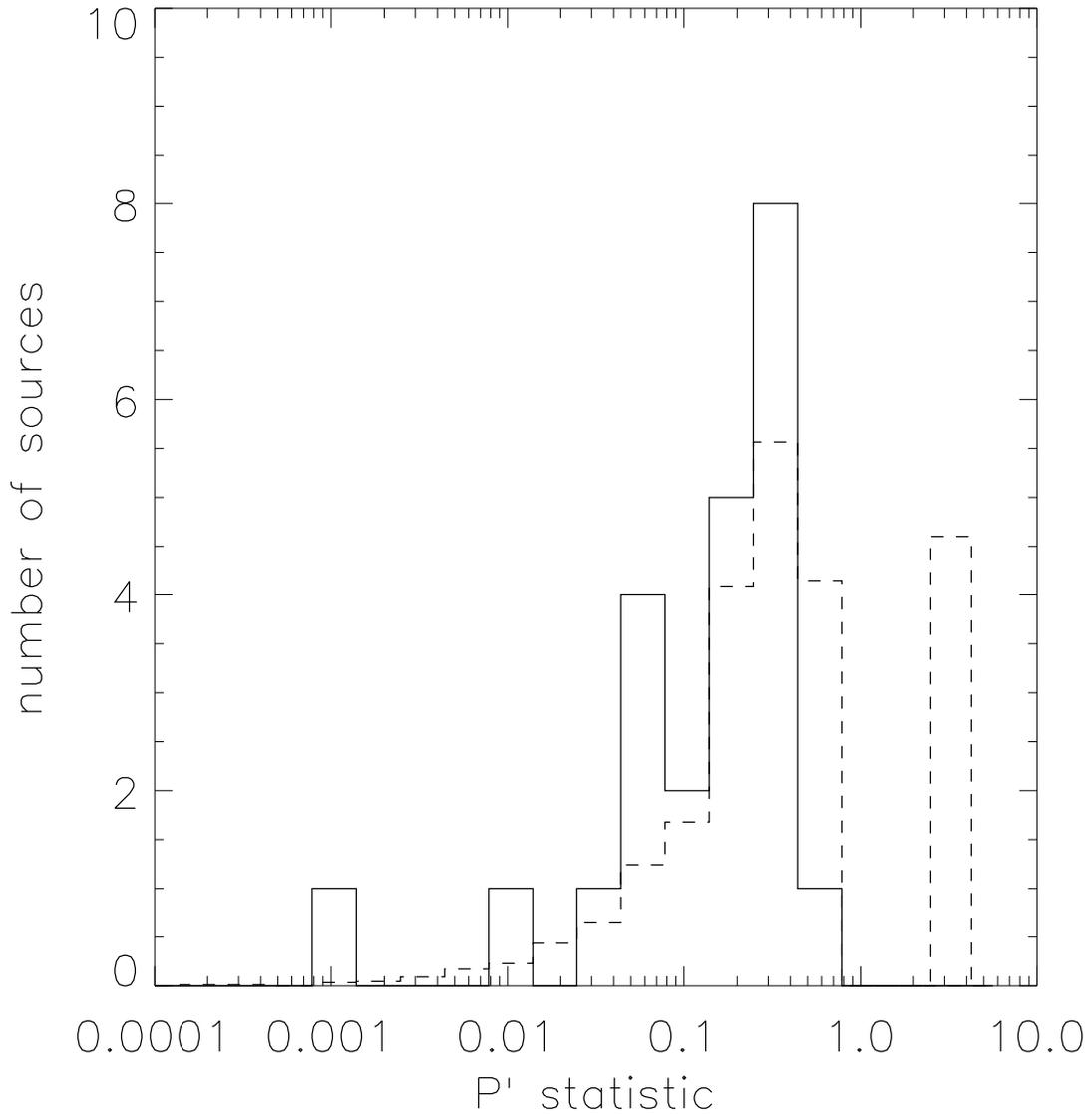}
\figcaption{The distribution of \pp \ values for the best near-IR selected identification (solid line).  
Overlaid (dashed line) are \pp \ values found when the 
identification  algorithm is run with random positions in place of SCUBA positions.  The single bin above 
\pp=1  contains the number of empty 
fields found  with the random positions.  In the real data there are no empty fields. This figure illustrates the fraction of  real identifications, as a function of \pp, that are expected to be spurious.   \label{pplot} }
\end{figure}

In Figure \ref{pplot}  we plot a histogram of the \pp \ values 
for the SCUBA  identifications (solid line) and for the Monte-Carlo simulations of our identification procedure (dashed 
line).   An excess of real identifications with \pp $\leq$ 0.1 over the number predicted by the Monte-Carlo simulation is apparent.  
Statistically we  expect two random associations at this level of significance but our procedure has selected seven. This implies that on order five of these seven sources are correctly identified.

One might expect that the addition of radio information to the NIR data would solidify a number of 
ambiguous identifications, but we find this is 
not the case.  Of the seven NIR-selected identifications  with $P^\prime <$ 0.1,  five of 
them are the same identifications selected by radio detections.   That is, {\it 
given only the NIR data our identification procedure securely identifies the same objects that are identified 
using radio data}.  Therefore, at these 
flux limits (14 $\mu$Jy) the radio data do not greatly improve the identifications over the NIR data. In particular it 
does not identify  objects 
missed by the NIR algorithm.  

Between \pp $\sim$ 0.1 and \pp $\sim$ 0.5  there 
is still an excess over random but it is less pronounced, only $\sim$ 3 sources.  Thus, in this region it is 
reasonable to expect a number of correct 
identifications but there are at least as many, and probably more,  random associations.   Moreover, it is impossible to know 
which specific identifications are correct 
simply based on the $P^\prime$ value.

\begin{figure}
\epsscale{1.0}
\plottwo{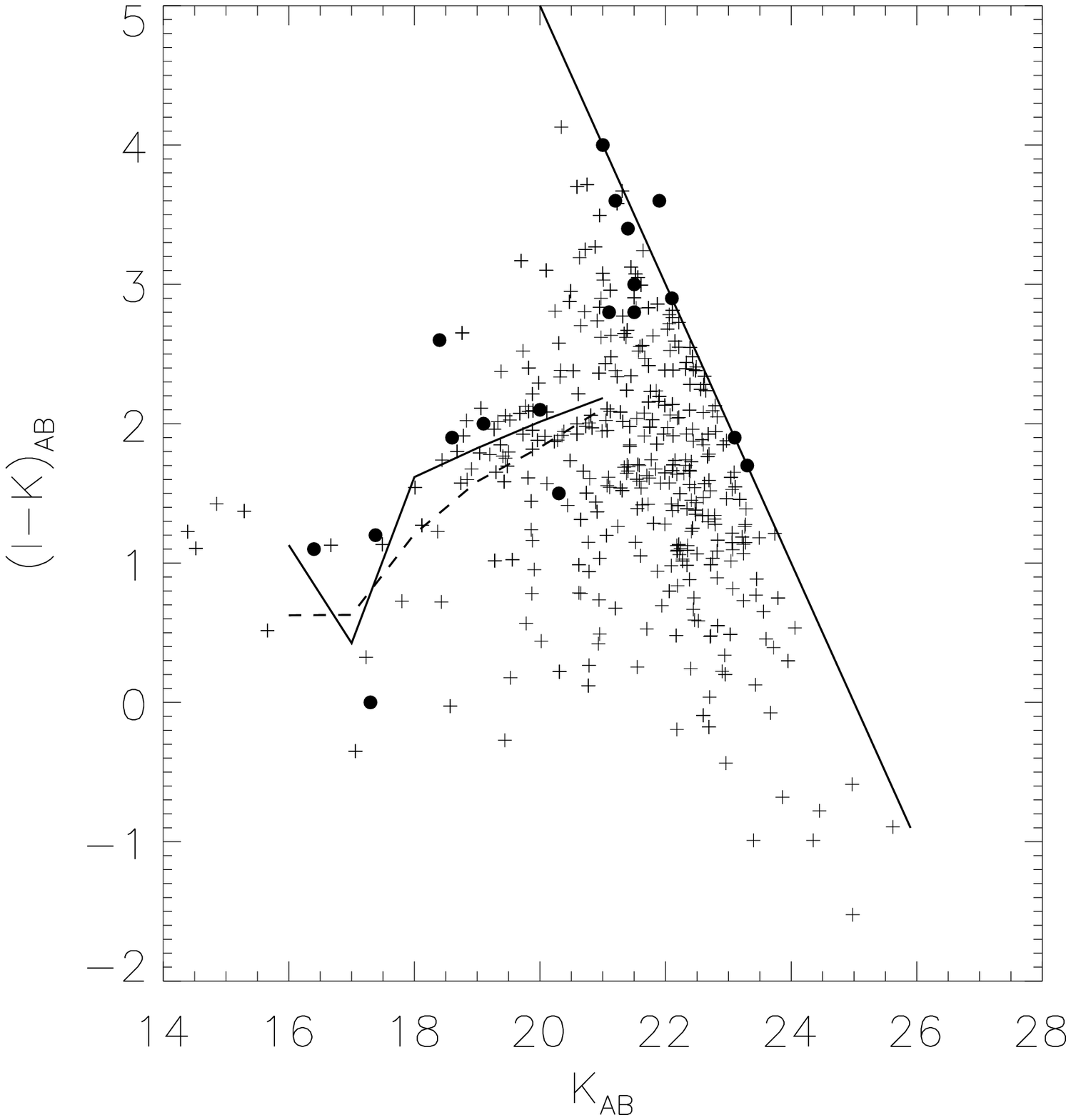}{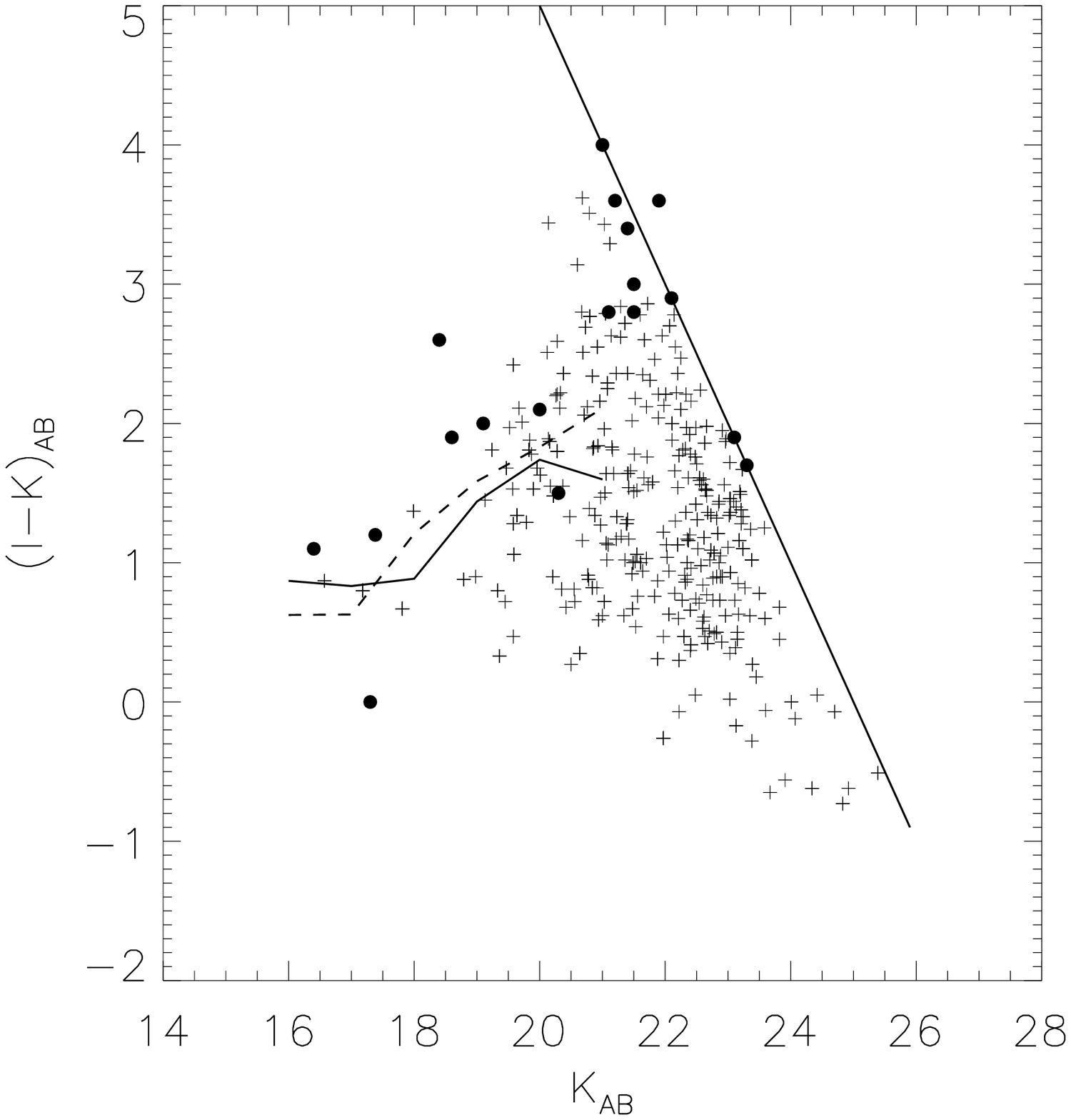}
\figcaption{The left plot shows the magnitudes and colors of the ``fake'' identifications selected when the 
identification algorithm is run with 
random positions in place of actual SCUBA positions (plus signs).  The right plot is the same but the identification 
algorithm has been modified to 
preferentially select  blue galaxies rather than  red galaxies. Overlaid on both plots is the mean color with magnitude for the entire 14\h field $K$ image (dashed line), and the mean colors for the fake identifications in each plot (solid line). For comparison, the statistically secure SCUBA identifications are shown by the solid points. \label{colmag_n}}
\end{figure}

 One might worry that   our identification algorithm will be overly biased toward red identifications but this appears to have a negligible effect. In Figure \ref{colmag_n} we show the magnitudes and colors of the identifications selected when the algorithm is run with random positions in place of SCUBA positions.   In this case the best identifications posses a 
wide range of magnitudes and 
colors  and are not concentrated in the region in which the good SCUBA 
identifications are located. There is a small offset towards redder average color for a given magnitude but this is not large enough to account for the unusually red colors of the statistically secure identifications.     Also 
shown in Figure \ref{colmag_n} are the colors and magnitudes of the Monte Carlo identifications 
found when we modify our 
identification algorithm to be biased toward blue objects, rather than red.  Again, the colors of the selected identifications span a 
wide range of values and are only offset by a small amount from the color of the general field population.

\subsection{Lyman-break galaxy identifications}

This field has also been surveyed for Lyman-break galaxies (LBGs) by two groups: Steidel and 
collaborators (personal communication) and the Canada-France-Deep-Fields Survey (CFDF) (\citet{mcc01}; Foucaud et al., in preparation).  The 
statistical submillimeter properties of the LBG population within the CUDSS fields  has been explored by \citet{web03a}, 
who 
also discuss the two LBG samples in 
detail.  In this work we will limit the discussion to  SCUBA sources with possible LBG counterparts. 

As with red galaxies and radio sources LBGs  have low 
surface densities 
and are therefore less likely to be randomly associated with SCUBA sources.
There are five LBGs within 8\arcsec \ of a SCUBA source and these are listed in Table \ref{lbgid}. However, one of these is within the error radius of source 14.9 and,  as 14.9 is securely identified with 
a 
radio source, the LBG is clearly a positional coincidence. 

The remaining LBG identifications are not secure for the following reasons. Given the $P$ values of the CFDF  LBG identifications  we would expect about  one random association. There 
are 
two (sources 14.6 and   
14.9) and therefore based purely on this one could conclude that 14.6 is properly identified with an LBG.  
However, as this is only an 
excess of one object over that expected randomly it is not statistically significant.  The $P$ values of the 
Steidel et al. identifications are 
all too high to be considered secure.  At these levels of significance we would expect two such identifications randomly and we 
have three.  Again, this is an 
excess of only one object of the number expected randomly and well within the shot noise.

  To 
estimate the $P$ statistic for these objects the surface density of LBGs over the entire survey field was 
used but, in fact, these four
objects are found amongst the densest concentration of LBGs in this field.   This could have a number of 
interpretations.  
It is most likely these identifications are simply chance coincidences occurring with greater frequency 
because of the higher surface density 
of objects in this region.  However, submillimeter sources have been associated with regions of optical 
over-densities \citep{cha01} and 
indeed, this small area within the survey field also contains the highest concentration of SCUBA sources in 
our survey.  Perhaps more 
LBGs in this region are submillimeter bright because of high-density environment effects, or this area  is an 
over-dense region in real-space 
where the SCUBA sources and LBGs are clustered together but not physically the same objects \citep{web03a}.  
  \

\subsection{The Positional Offsets}

\begin{figure}
\epsscale{1.0}
\plotone{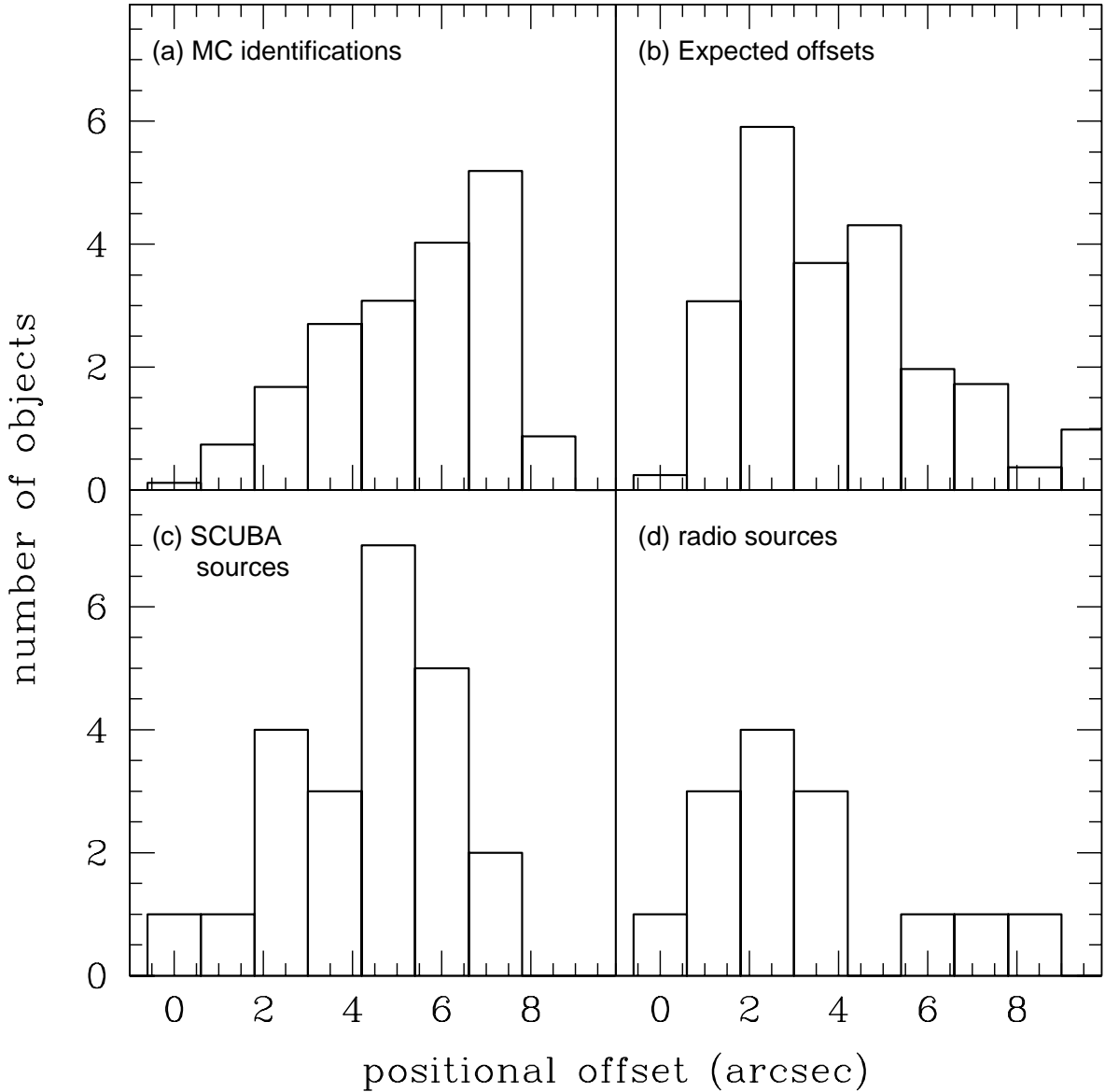}
\figcaption{In this figure we show four relevant offset distributions which are discussed in \S 3.5.  (a): The offsets found in the Monte-Carlo simulations of our identification procedure. (b): The distribution of offsets expected between the measured and true positions in the submillimeter data, due to noise and confusion.  (c): The offsets found between the submillimeter position and the NIR-selected identifications. (d): Same as (c) but considering only the radio detections.   \label{off}}
\end{figure}

In this section we consider four positional offset distributions of interest.  These distributions are shown in Figure \ref{off} and discussed below.

\begin{enumerate}

\item{Figure \ref{off}(a). The offsets derived in the Monte-Carlo simulations of the identification procedure.  These are the  offsets between the random positions chosen from the  $K$-image and the best ``fake'' identification chosen for each random position. This is the distribution that the incorrect identifications will follow.   These are not  uniformly distributed, but  peak at 7\arcsec, with very few at low offsets. }

\item{Figure \ref{off}(b). The  distribution of offsets between the measured and true positions of objects in the submillimeter data  due to the effects of noise and confusion.  These are the offsets that the real SCUBA sources and their true counterparts would be expected to follow.  As outlined in \S3.1 and Paper IV, this was estimated through simulations of the submillimeter data. This distribution peaks at $\sim$2-3\arcsec \ with an extended tail to large offsets.}

\item{Figure \ref{off}(c). The offsets between the submillimeter positions of the real SCUBA sources and their best NIR-selected identification. Clearly this distribution will be a superposition of the above two distributions, (a,b), as the identifications are a mixture of true counterparts and incorrect identifications. Indeed, if one compares Figure \ref{off}(c) with Figure \ref{off}(a), an excess of offsets at $<$ 5\arcsec \ is evident. The magnitude of this excess is roughly equal to the excess number of objects over random found by the NIR identification algorithm  (Figure \ref{pplot}).  }

\item{Figure \ref{off}(d). The offsets between the submillimeter positions and the radio positions  for the SCUBA sources detected in the radio.  These are secure identifications and thus these offsets  mirror the offsets of (2). To clearly show the distribution we have included the radio sources from both CUDSS fields (see Paper VI).   }
\end{enumerate}

\section{Notes On Individual Sources}
Below we discuss the properties of  individual identifications.  
Based on the reasoning discussed in the previous sections we have divided our sample into three 
identification categories: (1) those with radio detections (all but one are considered secure), 
(2) those with plausible identifications and, (3) those with  ambiguous or poor identifications.

\begin{figure}
\plotone{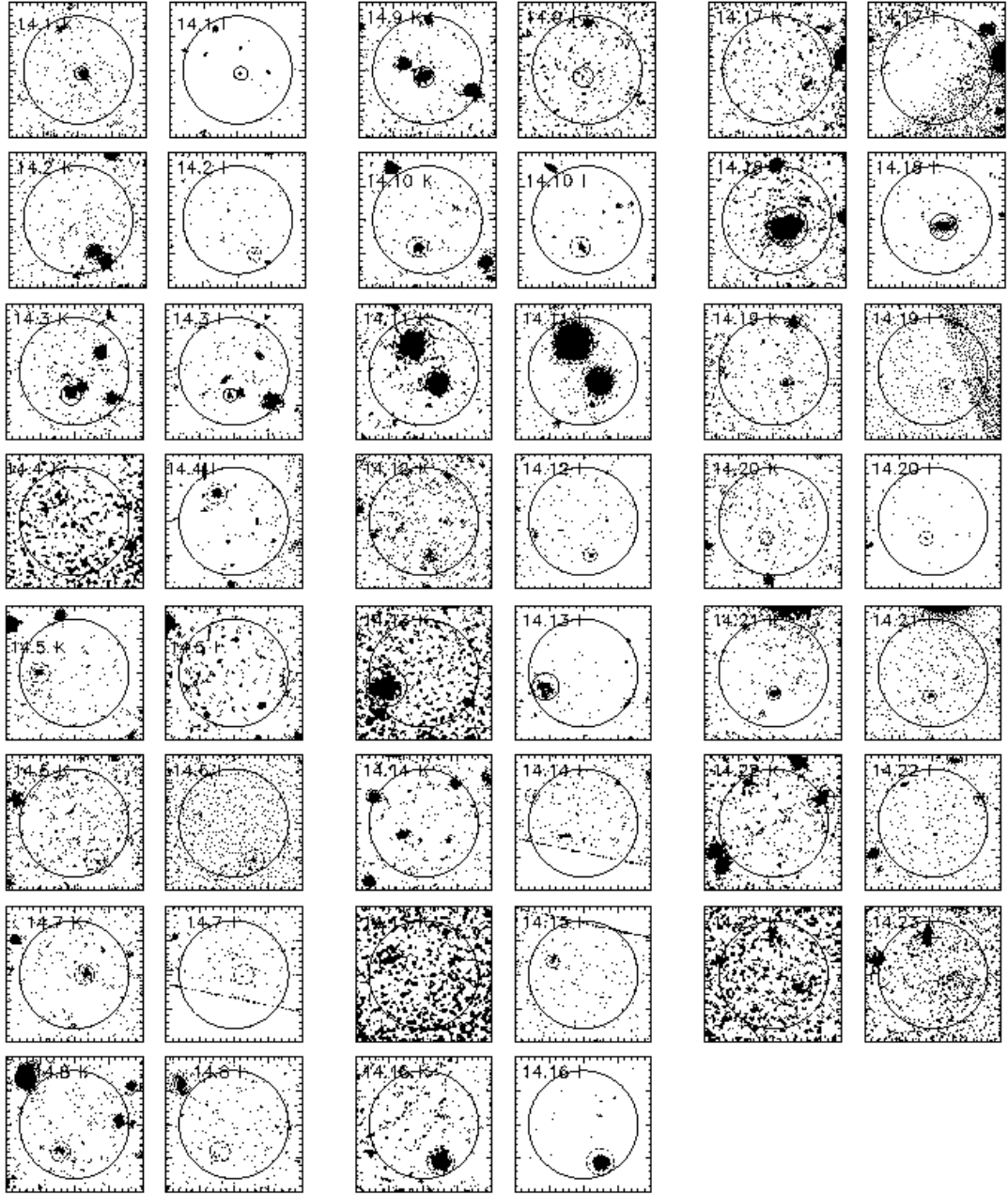}
\figcaption{The  $K$ (left of pair) and $I$ (right of pair)  images (20\arcsec$\times$20\arcsec) of all 23 sources in our 14\h catalogue, centered on the submillimeter position. The large circle  denotes the 8\arcsec search radius, or 90-95\% confidence region. The smaller circles mark the best NIR-selected identification for each source.  Solid circles correspond the objects which have also been detected at 1.4GHz, while dashed circles mark non-radio sources.  North is up and east is to the left. \label{postages}   }  
\end{figure}

In the following discussion we  define an ERO as objects with $(I-K)_{AB} \geq 
$ 2.7 \footnote{$(I-K)_{vega} \geq$ 4.0} and a VRO as objects with $(I-K)_{AB} \geq 
$ 2.2 \footnote{$(I-K)_{vega} \geq$ 3.5}.
Postage stamp images (20\arcsec $\times$ 20\arcsec)  in $I$ and $K$ of the  SCUBA sources  are shown in Figure \ref{postages}, and  HST $I$-band images of the four radio identifications with HST imaging are shown in Figure \ref{hst}.

We also discuss the redshift or redshift constraints for many of the sources.  When, as in most cases, spectroscopic measurements are not available we have estimated the redshift using the FIR and radio data (\citet{yun02}, hereafter referred to as the YC method) and NIR/optical colors (see \S 5).  \citet{yun02} used the entire radio-to-FIR SED to derive a photometric redshift and should therefore be a superior method to the 1.4GHz-to-850$\mu$m flux ratio method which only uses two data points (but see also \citet{bla03} for caveats).  The assumed template SED was produced by Yun \& Carilli from the observations of 23 local dusty starburst galaxies which span an order of magnitude in SFR and $L_{FIR}$.

\subsection{The Radio Sources}

\paragraph{CUDSS 14.1}

\begin{figure}
\plotone{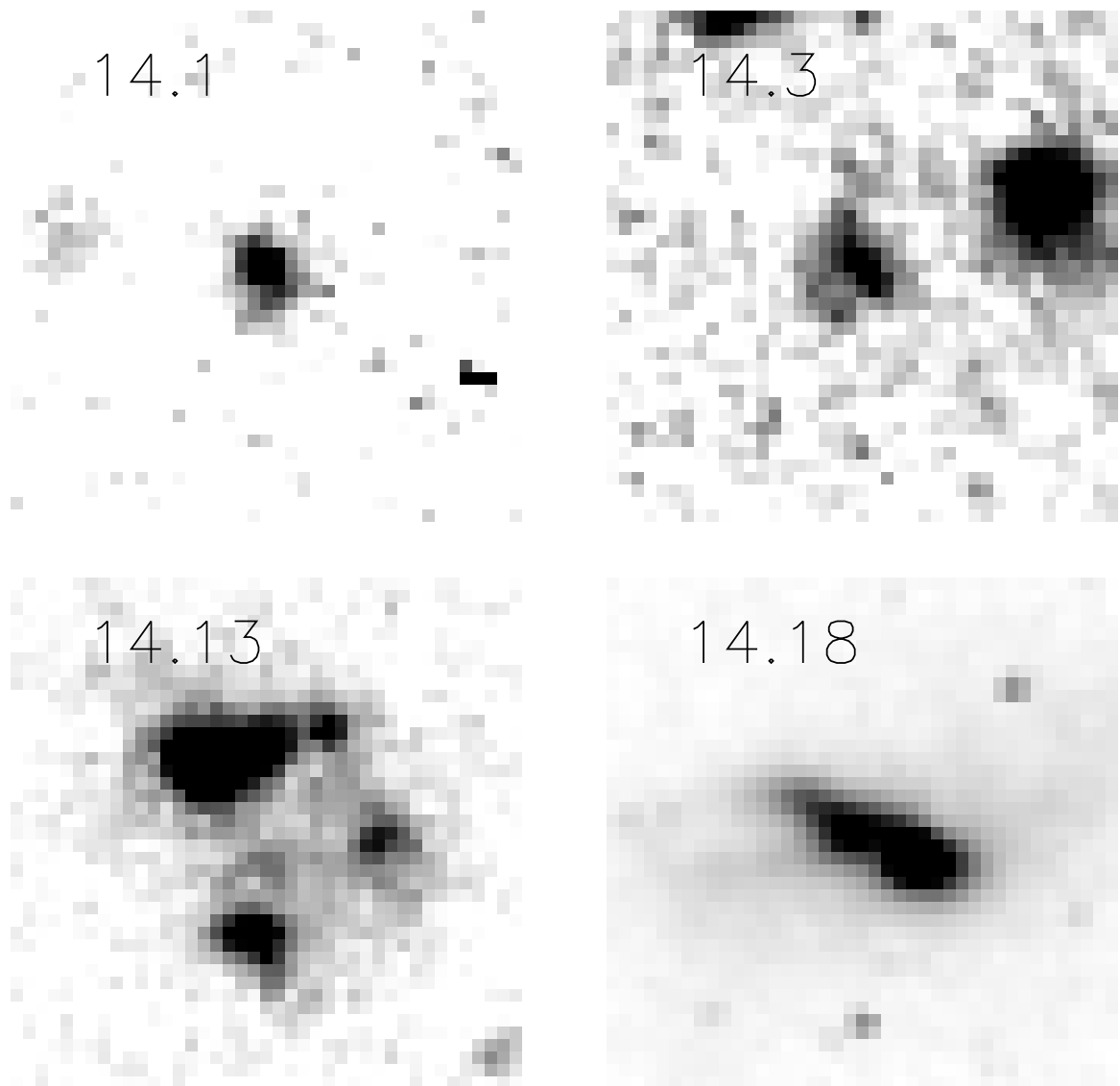}
\figcaption{HST $I$-band images of the four radio sources with HST imaging.  The images are 4$\times$4 
\arcsec. North is up and east is to the left. \label{hst} }
\end{figure}

 This  is the brightest object in our 14\h  field sample has been  extensively studied  at 1.4 GHz, 5 GHz, 1.3 mm,  optical ($VI$) and near-IR ($JHK$) wavelengths \citep{fom91,gea00}. It is classified as an ERO  with with  $(I-K)_{AB}$ = 3.6. The  HST $I$-band image shows asymmetrical morphology (Figure \ref{hst}). 

\begin{figure}
\epsscale{0.9}
\plottwo{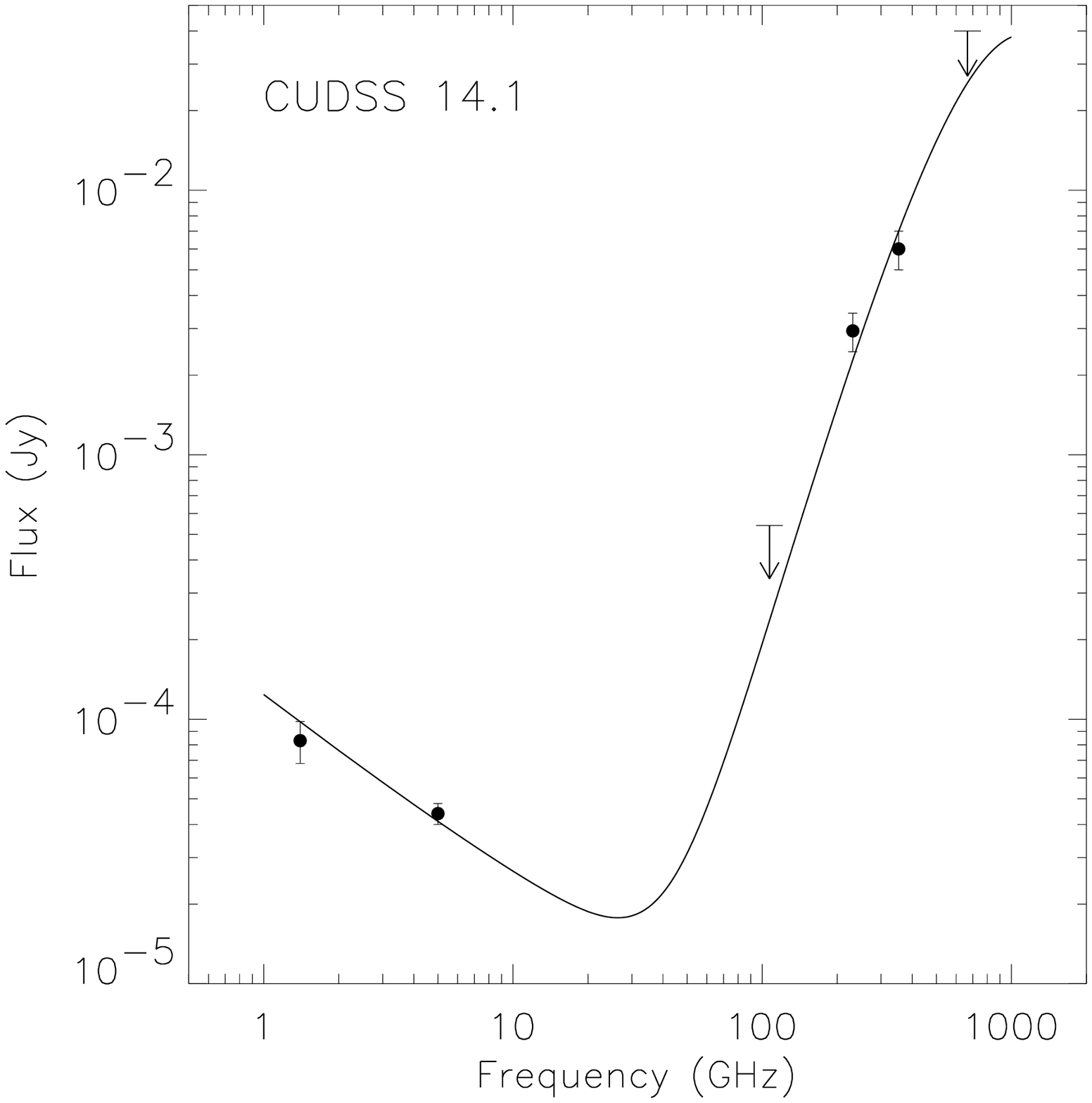}{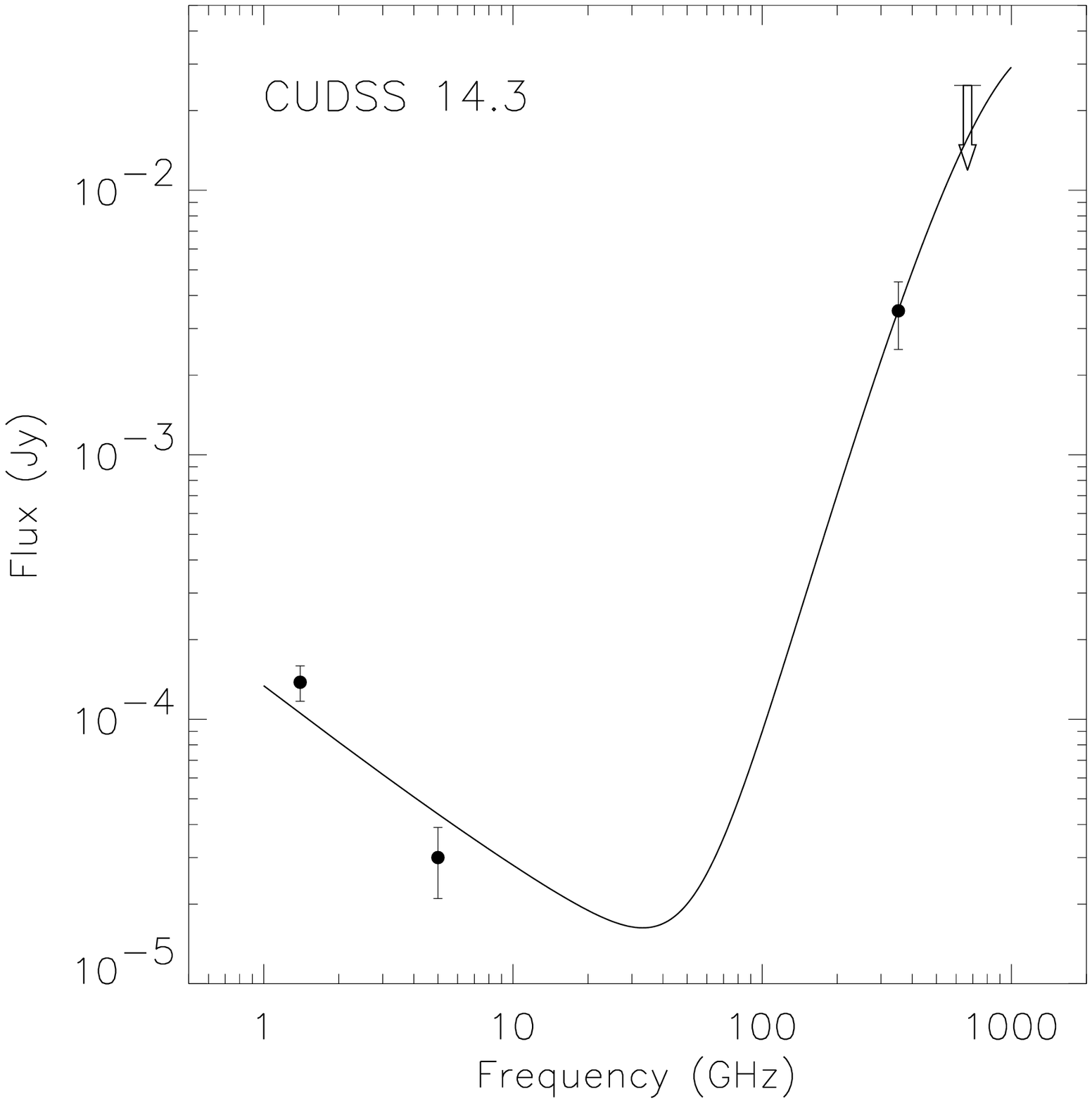}
\epsscale{1.9}
\plottwo{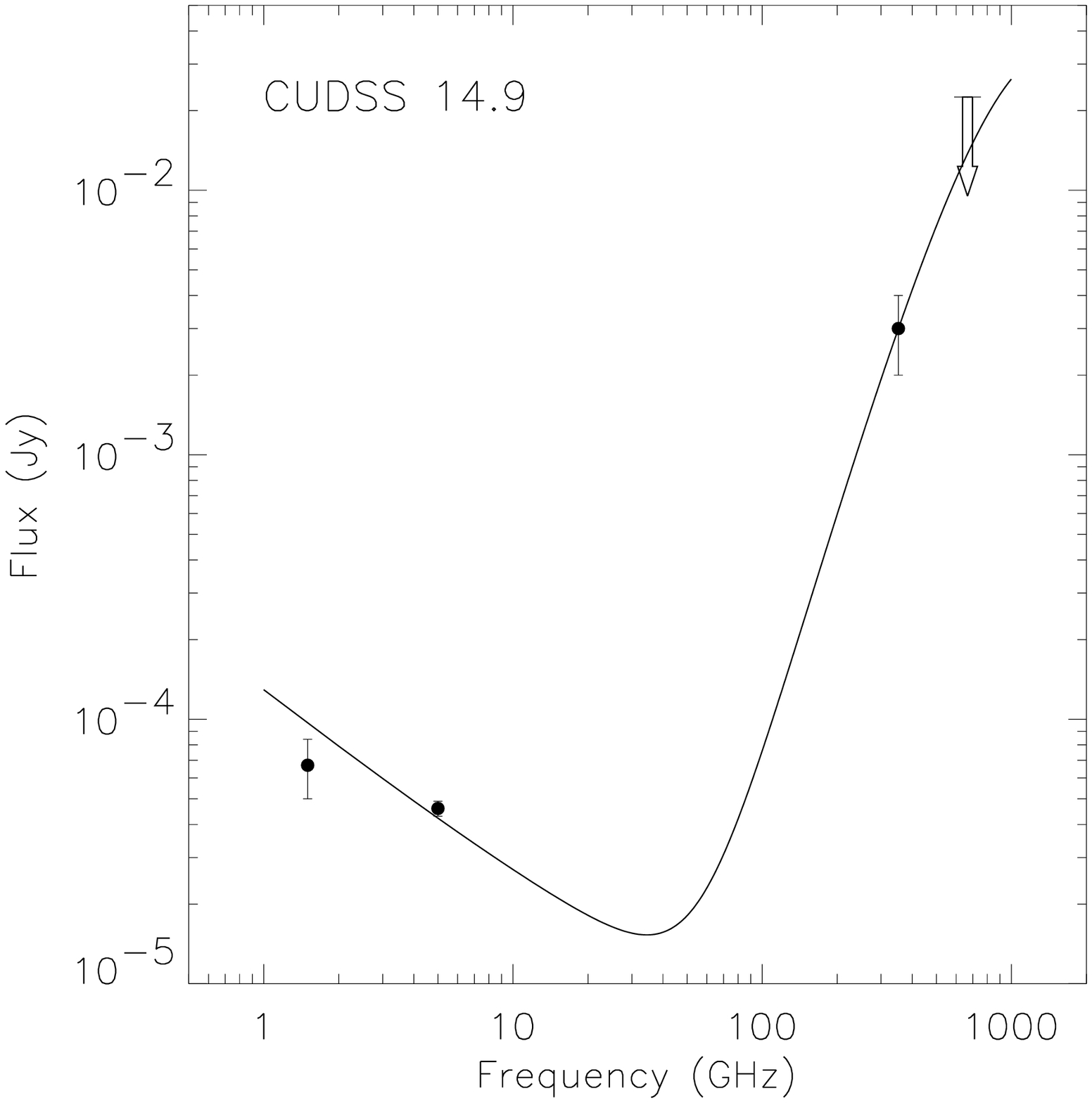}{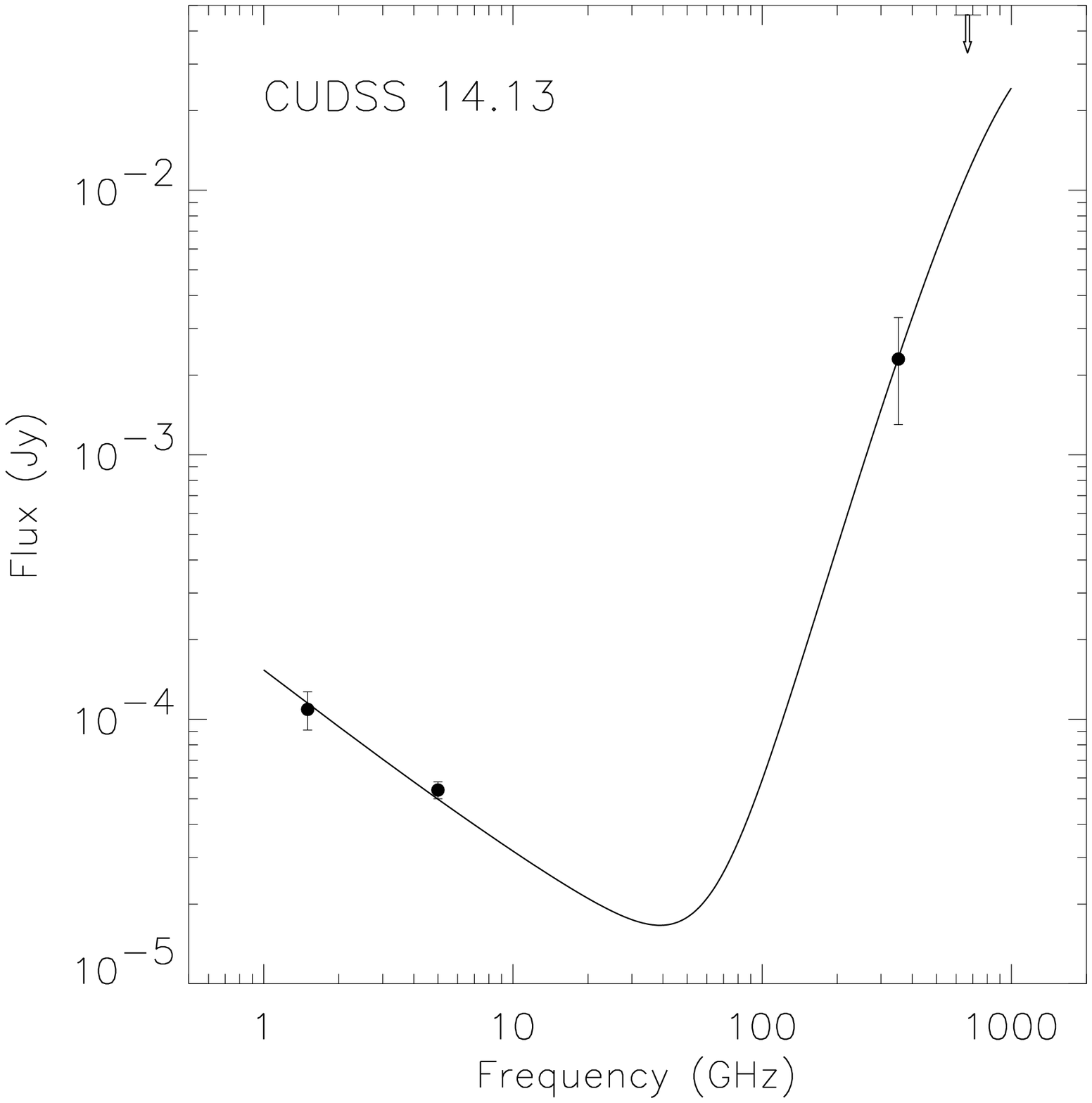}
\plotone{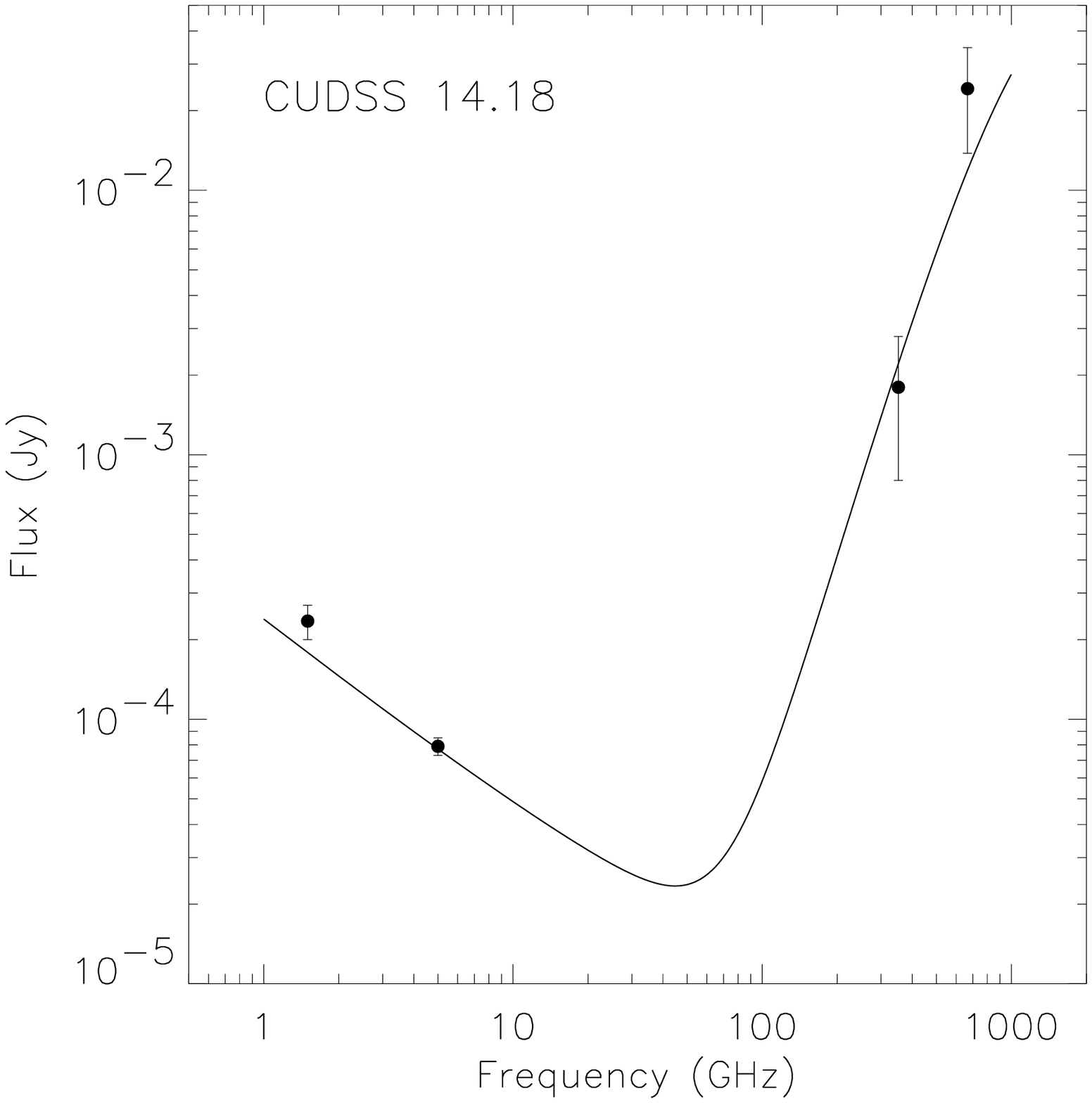}
\caption{ The YC 
templates (with  their best fit SFR and redshift) for the CUDSS+14 sources  plotted with the 
radio/submillimeter/FIR flux measurements 
for each source. \label{yc}}

\end{figure}

Using the YC method we estimate a redshift of $z$=2.3 $\pm$ 0.5 with a star formation rate of 635 M$_\odot$/year or $L_{FIR}$ = 3.7$\times$10$^{12}L_{\odot}$ \citep{ken98} (Figure \ref{yc}).  This is in good agreement with earlier estimates of the redshift \citep{gea00}. Assuming this source has a similar optical/NIR SED to source 14.13 (which is also well studied and has a spectroscopically determined redshift) places it in the range $z\sim$ 2.4-3.7.

\paragraph{CUDSS 14.3}
This object is identified with a radio source 15V23 \citep{fom91}.  It has a color of $(I-K)_{AB}$=2.4 and is classified as a VRO.
Though the object is faint, the HST $I$-band image clearly shows asymmetrical morphology (Figure \ref{hst}).

Following the YC method we estimate a redshift of $z$=1.7 $\pm$ 0.5 with a star formation rate of 315 M$_\odot$/year (Figure \ref{yc}) or $L_{FIR}$ = 1.8$\times$10$^{12}L_{\odot}$ \citep{ken98}.  Using source 14.18 as a NIR template  constrains this source to $z\sim$ 1.3-2.5.

\paragraph{CUDSS 14.9}

This object is securely identified with  radio source 15V67 \citep{fom91} with $(I-K)\gtsim$ 4.0 (it is undetected in the CFRS 
$I$-band image and no HST imaging exists in 
this region).  It is one of a chain of three extremely red objects ($(I-K)_{AB}$=3.48  and $(I-K)_{AB}\gtsim$4.1), that is 10\arcsec \ long. The YC redshift estimate is $z$=1.6 $\pm$ 0.6, and a star formation rate of 265 M$_\odot$/year  or $L_{FIR}$ = 1.5$\times$10$^{12}L_{\odot}$ \citep{ken98} (Figure \ref{yc}).

\paragraph{CUDSS 14.13}
This source is securely identified with radio source 15V23 \citep{fom91} also known as ISO 0 \citep{flo99a} and  CFRS 14.1157 \citep{ham95b}.  It has 
$(I-K)_{AB}$=2.6 and is classified as a VRO.  The  HST imaging 
shows an extended object ($\sim$2.5\arcsec) 
with multiple components separated by diffuse emission (Figure \ref{hst}).

 This  object has a spectroscopic 
redshift (Hector Flores, personal communication) of  $z$=1.15.  Using the YC method we find $z$=1.3 $\pm$ 0.4 with a star formation rate of 195 M$_\odot$/year or $L_{FIR}$ = 1.1$\times$10$^{12}L_{\odot}$ \citep{ken98}(Figure \ref{yc}).  This source has recently been detected at x-ray wavelengths \citep{was03}, and  likely contains an AGN. 

\paragraph{CUDSS 14.18}

This is the faintest object (at 850$\mu$m) in this catalogue and is  located in the deepest region of our map.  It is identified with 
radio source 15V24 \citep{fom91} also known as  ISO 5 \citep{flo99a} and  CFRS 14.1139 \citep{lil95a}.   HST imaging  shows 
disturbed morphology (Figure \ref{hst}).

 It has a spectroscopic redshift of $z$=0.66 \citep{lil95b}.  The YC redshift estimate places this objects at $z$=1.0$\pm$ 0.3 and a star formation rate of 165 M$_\odot$/year  or $L_{FIR}$ = 1.0$\times$10$^{12}L_{\odot}$ \citep{ken98}.

\paragraph{CUDSS 14.19}
This source has a radio detection at 8.5{\arcsec}, just beyond the search radius.  We expect 5-10\% of the sample to have identifications that lie beyond 8{\arcsec} and so this identification is possible but by no means secure.  A second possible identification is found with the NIR identification algorithm, which lies at a smaller offset of 3{\arcsec} but is not statistically significant.

\subsection{Plausible Identifications}

The sources discussed in this section do not have radio detections. However, we know that a number of the non-radio sources have been correctly identified (\S 3) though it is not possible to know which ones simply based on their \pp \ values. In the following we discuss three sets of identifications:

\begin{enumerate}

\item{Sources with NIR-determined \pp$\leq$ 0.1 which do not have radio detections: sources 14.23 and 14.11.  As discussed in \S 3 we expect $\sim$ 2 identifications due to positional coincidence at this level of significance.  However, since we are dealing with small number statistics it is possible that none or both of these identifications are correct.}

\paragraph{CUDSS 14.11}

This object is best identified with a very bright object ($I_{AB}$=17.3) with $(I-K)_{AB}$ = -0.05.     The lack of a  radio  detection 
for this source constrains its redshift to $z>$ 1.1.  However, the optical identification has colors 
and morphology consistent with a 
low-redshift elliptical and this is, therefore, unlikely to be the correct identification. An alternative explanation (see next section), which is consistent with the radio-estimated redshift lower-limit,  is that it is a foreground gravitational lens which is amplifying a  submillimeter-bright object at higher redshift.

\paragraph{CUDSS 14.23}

This object is identified with an ERO, with $(I-K)_{AB}$=3.6.   As we discuss in \S 5, this source as NIR colors consistent with the radio detected objects if placed at redshifts of 2.5-3.7, and therefore this identification is considered likely.

\item{Sources with LBG associations:  These identifications were outlined in \S 3.4, and we concluded that none of the 4 LBG identifications could be considered secure. We discuss only one identification of interest.}

\paragraph{CUDSS 14.7}
The best identification for this object is the Steidel et al. LBG West2 MMD13, which is a spectroscopically 
confirmed QSO at $z$=2.913.  It lies in 
the single most over-dense region of the Steidel et al. 14-hour LBG survey.   There is a second possible identification in this field which 
at a comparable level of significance.   
This galaxy, at an offset of 2.1\arcsec, has an $(I-K)_{AB}\gtsim$ 2.8, thus is an ERO.  
Because there is no $I$ detection we only 
have an lower limit on the color and therefore an upper limit on the \pp \ value  and  may 
therefore be of higher significance than the 
LBG.   

\item{Sources with NIR-determined \pp$>$ 0.1: In \S 3.3 we claimed that there was an excess of 3 identifications at this level of significance, over that expected randomly, and therefore at least this number of identifications are correct.  Again, it is not possible to identify which sources have been correctly identified based solely on their \pp \ values.  Below we discuss those objects with interesting properties such as extremely red colors.} 

\paragraph{CUDSS 14.2}
 The best identification for this object is a very red galaxy  5.6 \arcsec \
away from the submillimeter position.  Two arcseconds away from this identification (7.8\arcsec \ from the 
submillimeter position) there is a second 
red object.  These two objects are both classified as EROs and have  colors of $(I-K)_{AB}$=3.5 and 
$(I-K)_{AB}$=3.7 respectively. 

\paragraph{CUDSS 14.5}

This object is best identified with a faint ($K_{AB}$=22.3) ERO ($(I-K)_{AB}$=2.7) at an offset of 5\arcsec. It is just detected in the CFRS $I$-band image and unfortunately, no HST data exists in this region.

\paragraph{CUDSS 14.17}
This source is identified with ISO 195 \citep{flo99a}  which lies at an offset of 10.3\arcsec. Though 
larger than our nominal search radius we do 
expect 1-2 sources to have identifications with offsets from the submillimeter position of $>$ 8\arcsec.   
However,  at such large distances the number of 
random coincidences rises significantly.    This object has no spectroscopic redshift and cannot be 
ruled out by the radio redshift lower limit of 
$z>$ 1.6.  

\end{enumerate}

\subsection{Gravitational lensing effects}

The possibility that a significant fraction of SCUBA sources in blank-field surveys could be lensed by foreground objects was first discussed by \citet{bla96} before the commissioning of SCUBA.  Since then, a number of SCUBA sources that were (or would be)  identified with relatively nondescript  low-redshift galaxies, based on positional coincidence arguments, have, upon further study, turned out to be more distant objects, possibly magnified by the foreground object which was originally thought to be the identification \citep{sma99,dunl02,cha02}. 

In our own survey there are a number of sources whose best identification is with a low-redshift system ($z<$ 1), two in the 14\h field (14.11 and 14.18) and four in the 3\h field (3.8, 3.10, 3.15, 3.25; \citet{web03b}).  Since none of these identifications have been verified through the detection of CO it is possible that all six are incorrect. If so,  their close proximity to the SCUBA positions  may  be the result of lensing. However, the available evidence suggests this is not the case but rather that we have located the correct identification for all but one of these objects.

These systems, with the with the exception of source 3.25 (which is faint, small and diffuse, and source 14.11, which is bright, large, and spheroidal) exhibit  unusual morphology suggestive of merger activity.  
Five of the six sources (all but 14.11)  have detections at 1.4GHz, coincident with the optical identification, and  four of these (all but 3.25) have also been detected at 15$\mu$m.  These detections  strengthen the statistical argument that  these objects have been correctly identified since the surface density of radio and ISO populations is relatively low.  More importantly, these are exactly the signatures we would expect for lower redshift objects. 

However, this doesn't remove the possibility that the 1.4GHz and 15$\mu$m flux has  been lensed and the geometry of the lens system is such that the observed  offsets between the source and lens are within the astrometric errors.  However, if this were the case we would  expect the 1.4GHz-to-850$\mu$m flux ratio to yield a redshift estimate consistent with a distant background source.  In fact, all but sources 14.11 and 14.18, have  1.4GHz-to-850$\mu$m flux ratios consistent with a low-redshift ($z<$ 1) system. Source 14.18 is estimated to lie at $z\sim$ 1.4$\pm$ 0.3, which is not so large a discrepancy from the spectroscopic redshift of $z$=0.66 that it could be considered clear  evidence for lensing  (see for example \citet{dunl02}).  An alternative scenario is that although the submillimeter source has been amplified, the radio and ISO emission belong to the foreground galaxy.  In this case the SCUBA source could lie at very high-redshift but the  1.4GHz-to-850$\mu$m flux would produce an erroneously low-redshift. As outlined above (and in \S 3.2) this is possible but statistically unlikely.  Thus we conclude that sources 3.8, 3,10, 3,15, 3.25 and 14.18 have been correctly identified with low-redshift systems.

Source 14.11, on the other hand, is very suggestive of lensing.  Though the optical identification does not have a measured redshift its color, angular size, morphology,  and apparent magnitude are all consistent with a low-redshift elliptical. Not only is it difficult to reconcile strong submillimeter emission with this object, its lack of a radio detection places it at  $z>$ 1.1. It is likely, then, that this is not the correct identification. Whether or not it is actually a lensed system cannot be addressed by these data.

\section{Discussion}

The first part of this work was concerned with identifying the multi-wavelength counterparts of the sources in the 14\h field.  In the remaining sections we will discuss the general properties of the population and expand the discussion to include both the 14\h and 3\h catalogues.

There are a number of securely identified sources in both fields which have reliable redshift estimates and relatively well studied SEDs.  We will uses these objects as templates to draw tentative conclusions about the nature of the counterparts of all the objects. We will argue that we have correctly identified $\sim$ 50\% of the catalogue, and that the remaining unidentified sources are likely faint, red objects beyond the detection limits of this work.

\subsection{The 3\h Field Identifications}

  The radio and mid-IR identifications for the 3\h field sources were presented in Paper IV, and the optical/NIR counterparts are discussed in Clements et al., (in preparation). Clements et 
al. employ a slightly different algorithm for identifying counterparts 
than this work 
and use   $K$-band images using UFTI camera on the UKIRT telescope as  the primary NIR data set. In 
order to directly compare the 
identifications of the 3\h field sources  with those of the 14\h field we have repeated the above 14\h field identification analysis 
on the 3\h sources using the 
CFHTIR  data.  This analysis produces a list of candidate identifications in good agreement with those of 
Clements et al. 

There are   11  3\h field sources whose  NIR identifications have  \pp $\leq$ 0.1.
Statistically, we  expect about three of these to be due to chance coincidence and the remaining 
eight identifications to be correct.  Working from the radio data, there are 9 sources in the 3\h catalogue which have been detected at 1.4 GHz (11 mJy, 1$\sigma$), six of which are the same identifications as those selected by the NIR algorithm with \pp $\leq$ 0.1.  Of the remaining three radio identifications two are still the best identification selected by the algorithm, but at a lower level of significance.  The remaining source (3.17) is the only  radio-detected source in our catalogue which does  not  have an  optical/NIR counterpart to the depths of these data.
Thus, based on comparison to the radio identifications, our NIR algorithm remains successful at selecting  the correct counterparts  in the 3\h field. However, there may be a higher level of contamination (5/11) for the 3\h field than  for the 14\h field (2/7), (if these are indeed incorrect identifications), two sources are not identified with  confidence, and one is missed due to its  faintness.

\subsection{Properties of the identifications}

\begin{figure}
\epsscale{0.8}
\plotone{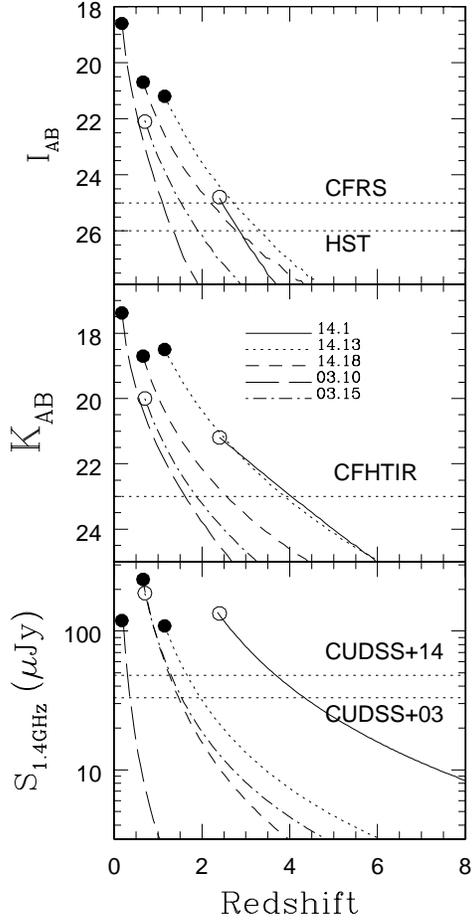}
\figcaption{The $I$,$K$ and 1.4 GHz fluxes as a function of redshift for sources 14.1 (blue), 14.13 
(magenta), 14.9 (cyan), 3.10 (red), and 3.15 (green). The solid points represent objects with spectroscopic redshifts.  Sources with open points have redshifts estimated through the YC method.  The $I$ and $K$ 
magnitudes as a function of redshift have been estimated directly from the optical/NIR data of the identifications.  The radio 
spectral indices were determined using the 
1.4 GHz data of Yun et al. (personal communication) and the 5 GHz data of  \citep{fom91}. Also shown 
are the detection limits at each wavelength of 
observation. 
\label{tracks}}

\end{figure}

In Figure \ref{tracks} we show  the $I$, $K$ and 1.4 GHz fluxes of five radio-detected objects  as they would 
appear when placed at higher redshifts. These are sources  14.1, 14.13, 14.18, 3.10, and 3.15 which have measurements in at least four  optical/NIR filters.  Sources 14.13, 14.18, and 3.10 have 
spectroscopically measured  redshifts, while for  14.1 and 3.15 we have assumed the radio-far-infrared 
redshift estimate.   For the 3\h field sources, which have only 1.4 GHz 
measurements, a spectral index typical 
of star forming galaxies was adopted: $S \propto {\nu}^{-0.5}$ \citep{fom91}.

A number of interesting conclusions may be drawn from this figure.  
Firstly, the optically faint natures of these objects do not  require them to lie at very 
high-redshift.  All of these objects reach $I_{AB}$ = 26 between $z \sim$ 1.3 (3.10) 
and $z \sim$ 3.5 (14.1).  
They are all visible at $K$ to higher redshifts than at $I$, but would drop out of our $K$ image between $z 
\sim$ 1.7 (3.10) and $z \sim$ 4.5 
(14.1 and 14.13).  

Secondly, all of these objects become undetectable at 1.4 GHz (to our 1.4 GHz limits)  at significantly 
lower redshift than they do at $K$.  
For example, source 14.1 would drop out of the radio survey at $z \sim$ 3.3, while remaining visible at $K$ 
to $z\sim$ 4.5.  Source 3.10 
becomes undetectable at 1.4 GHz at $z \sim$ 0.5, while still visible at $K$ to $z\sim$ 1.7.  

This has important consequences for the 
identifications.   If these radio sources are appropriate templates for the entire population there should be no 
submillimeter sources in the 
sample which are detected in the radio but have no $K$-band counterpart to our limits.  Indeed there are none in the 14\h field and only one in the 3\h field.  Moreover, there will be counterparts with similar optical/NIR properties to the radio-detected objects, but which have no radio detection.  We may use this fact to identify likely counterparts  in the absence of radio information.

\begin{figure}
\epsscale{1.0}
\plotone{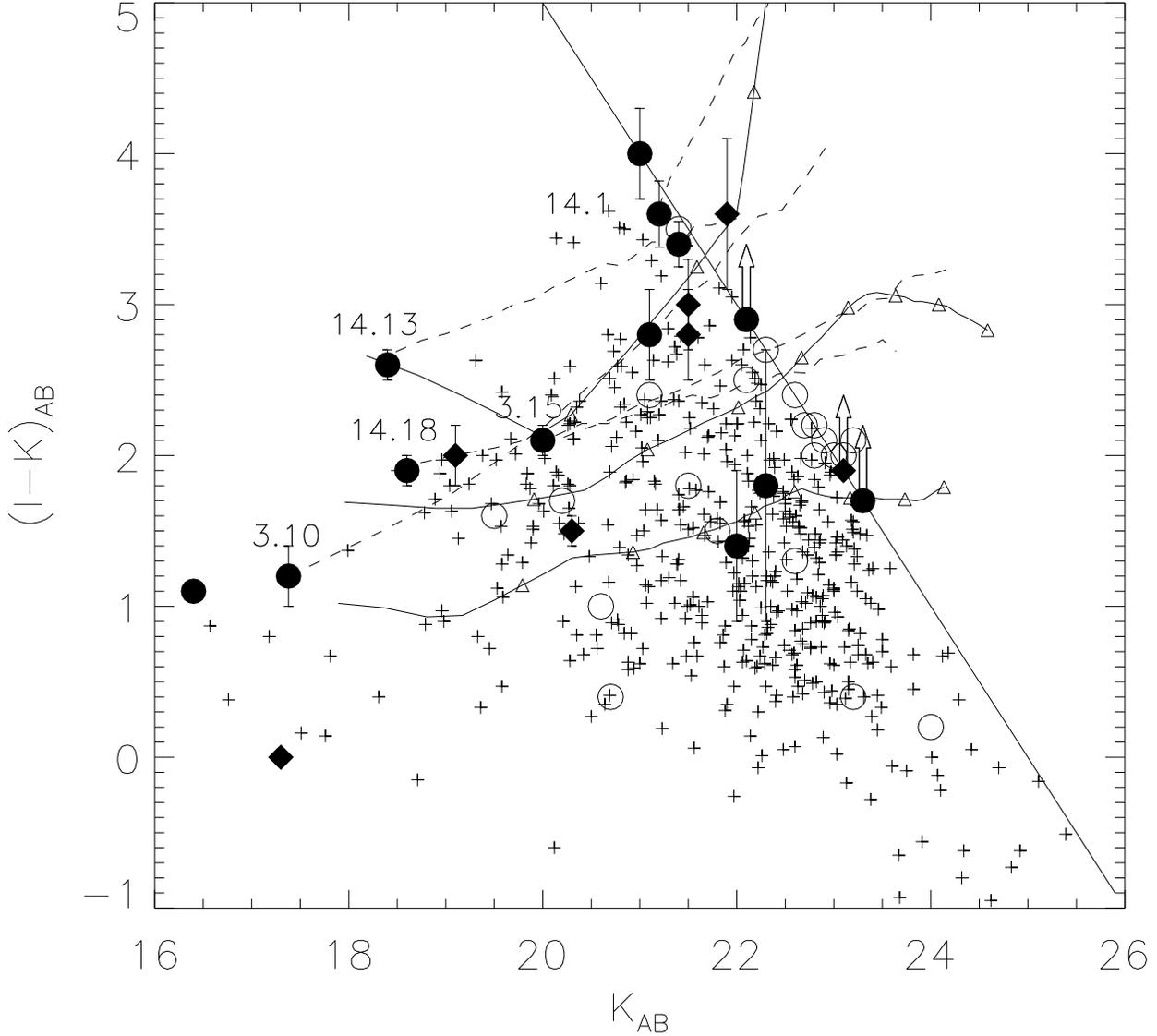}
\figcaption{The NIR color-magnitude diagram for the identifications in both the 14\h and 3\h fields (see Clements et al., in preparation).  The solid circles correspond to the radio-detected objects.  The solid diamonds denote those identifications with \pp $<$ 0.1 but without a radio detection.  The open circles show the best identification for the remaining objects (except for the empty fields). 
   Not included in this plot are the possible LBG identifications (whose colors are in the $grz$ filter system).  The solid, diagonal  black line denotes the $I$ detection limit 
of the CFRS.  The dashed 
lines are tracks of $(I-K)_{AB}$, $K_{AB}$ with redshift for sources 14.1, 14.3, 14.18, 3.10, and 3.15 
(see Figure \ref{tracks}).  The three  
solid  lines show the predicted colors for the three ULIRGs studied by \citet{tre99}.  These have been 
scaled to $M_{K_{AB}}$ = 
24.4 (approximately $M^{*}$-2).  The tracks begin at $z$=0.5 and are marked (small triangles) every 0.5 
step in redshift.  
\label{colmag}}
\end{figure}

Figure \ref{colmag} presents a NIR color-magnitude diagram for the best identification for each source in the 14\h  and 3\h catalogues.   The solid circles
correspond to the identifications with radio detections, and the solid diamonds to those with  \pp $\leq$ 0.1 but with no radio detection.  The open 
symbols correspond to the best 
identification for each of the remaining SCUBA sources.   For comparison the magnitudes and colors of 
all the galaxies in 
the CFHTIR 14\h  image  are also shown (plus signs). The bulk of identifications with radio detections or with  \pp$\leq$ 0.1  lie along the upper envelope of the general population, and 
are among the reddest objects at each magnitude

Overlaid on Figure \ref{colmag} are two sets of tracks.  The dashed lines show the extrapolated magnitudes 
and colors, as a function of 
redshift, for the  five  radio sources from Figure \ref{tracks}. 
Also shown (solid lines) are the predicted NIR colors and magnitudes for three ULIRGs studied 
by \citet{tre99} as they would 
appear at higher redshifts.  Though all three are ULIRGs they do not have consistent optical/NIR 
colors, particularly at 
high-redshift.  In particular, IRAS F12112+0305 (the bluest track) remains relatively blue at all redshifts. 
Thus it is not a foregone conclusion that these objects must have  very red optical/NIR colors because simply because they are dusty.
Indeed, sources 3.6, 3.25, and 3.27 are much bluer than the other radio sources (albeit with large uncertainties on their color) and lie along the track of  IRAS F12112+0305.

\subsection{Constraining the non-radio counterparts}

 Many of the non-radio identifications   lie along 
the extrapolated tracks of the radio counterparts Figure \ref{colmag}, indicating that 
they may be similar to the  securely identified sources, but lie at higher redshifts. 
We can use this information to identify likely counterparts and estimate redshifts for roughly 50\% of the sample.

We begin by discussing those objects with \pp $\leq$ 0.1 which do not have a radio detection.  In the 14\h field these are sources 14.11 and 14.23.   Source 14.23 lies within the same region as 
sources 14.1, 14.3, and 
14.9.  In fact, it has the exact color and magnitude as source 14.13 if placed at $z$ = 
3.1.  Source 14.11, on the 
other hand, lies in the lower left corner of the plot with the bright, blue galaxies, far away from the radio 
sources.  Based on this it is 
reasonable to conclude that source 14.11 is incorrectly identified while 14.23 is likely correct (see \S 4.3 for further arguments that 14.11 has not been correctly identified).  The fact that 
source 14.23 has not been 
detected at 1.4 GHz is not worrying as source 14.13 would not be detected in the radio beyond $z \sim$ 
1.2, though it would be 
visible at $I$ and $K$ to $z\sim$ 3.5 and $z\sim$ 4 respectively.  
 
In the 3\h field sources 3.2, 3.4, 3.5, 3.13, 3.16 have \pp $\leq$ 0.1 but do not have radio detections.   Three of these (sources 3.4, 3.13, and 3.22) have colors consistent with radio sources at higher redshift, and we therefore take these to be the correct counterparts.  

There are four sources (14.2, 14.5, 14.14, and 14.22) with  \pp$>$ 0.1 and  colors consistent with the radio identifications at higher redshifts. This is approximately the  number  of identifications expected over random associations at  this level of significance.

To summarize, given the radio and ISO identifications and those NIR selected identifications deemed correct based on their location in the color-magnitude diagram  we have identified likely counterparts for 50\% of the sources in our submillimeter catalogue. Later in the paper we will argue that the remaining 50\% are likely faint and red and lie below the detection limits of these data.

\subsection{Data constraints on the redshift distribution}

 Simple redshift estimates may be made for objects 
based on optical/NIR and colours using the 
secure, well-studied, low-redshift identifications as templates.  For each identification we find the  
template SED (Figure \ref{tracks}) and redshift that best 
describes the source's NIR magnitude and colour.  The resulting redshift estimates are listed in Table 
\ref{redshifts} along with the millimeteric and spectroscopic redshifts.

The mid-IR and submillimeter colours may also be used to roughly constrain redshifts.
 Four sources in our  catalogue have 450\micron \ detections and  eight sources are identified with 
15\micron \ ISO detections.  In Papers IV and VI   we 
argued that provided these objects have similar SEDs to typical local ULIRGs and dusty starburst galaxies these 
11 sources (one is detected at both 450\micron \
and 
15\micron) must lie at $z\la$3. Indeed, two of these are spectroscopically confirmed to lie at $z$=0.660 (14.18)
and $z$=1.15 (14.13).    

We do not have enough information to estimate a redshift distribution.  
However, assuming each source lies at its redshift lower-limit places a lower-limit on the median redshift 
of the population of 
$z_{med} \geq$  1.4.  At least 4\% of the sources must lie at $z\la$ 1.0 based on sources 3.10 and 14.18 
which are secure identifications with 
spectroscopic redshifts. 
If source 3.8 is correctly identified, and sources 3.15 and  3.25  have a correctly estimated redshift, the 
fraction of sources below $z\sim$ 1.0 rises to 10\%.  The remaining 45 sources are constrained 
to $z\geq$ 1.0 based on their non-detection at 1.4 GHz.

As our redshift estimates are limited to low-redshift it is difficult to assess the number of sources in our catalogue which lie at redshifts   $z \gtsim$ 2-3. Though we have possible identifications with a small number of  LBGs, these are not secure, and only one is spectroscopically confirmed to lie at $z\sim$ 3.
However,  based on the radio, ISO and 450\micron \ data we claim that no more than $\sim$ 60\% may lie at $z >$ 3.0. These results are in agreement with the general consensus in the literature that the bulk of the 
sources lie at $z >$ 1  with a median redshift $z \sim$ 2 - 3  \citep{bar99b,sma00,bar00,fox02,sma02,ivi02,hug03,are03}.


\subsection{Models of the Redshift Distribution}

In order to get a feel for the types of  redshift distributions that are consistent with the identifications of this paper, we have developed simple models of the data.  Using the radio galaxies (14.1, 14.13, 14.18, 3.10) as templates, we randomly populate Gaussian redshift distributions of varying peak redshifts and widths, and recover the $I$ and $K$ magnitudes of the counterparts.  As an aside, two caveats regarding sources 14.18 and sources 3.10 should be remembered in this analysis. Firstly, in \S4.3 we discssed the possibility that sources 14.18 and 3.10 were incorrectly identified with foreground, low-redshift lenses, and though we concluded that this was unlikely it may in fact be the case.  Secondly, even if these low-redshift objects are correctly identified, they may be unrepresentative of the higher-redshift population \citep{dun00b,cha02}.  However, as there  currently no strong evidence to suggest otherwise we proceed with the simple assumption that  these objects are not unique.

\begin{figure}
\plotone{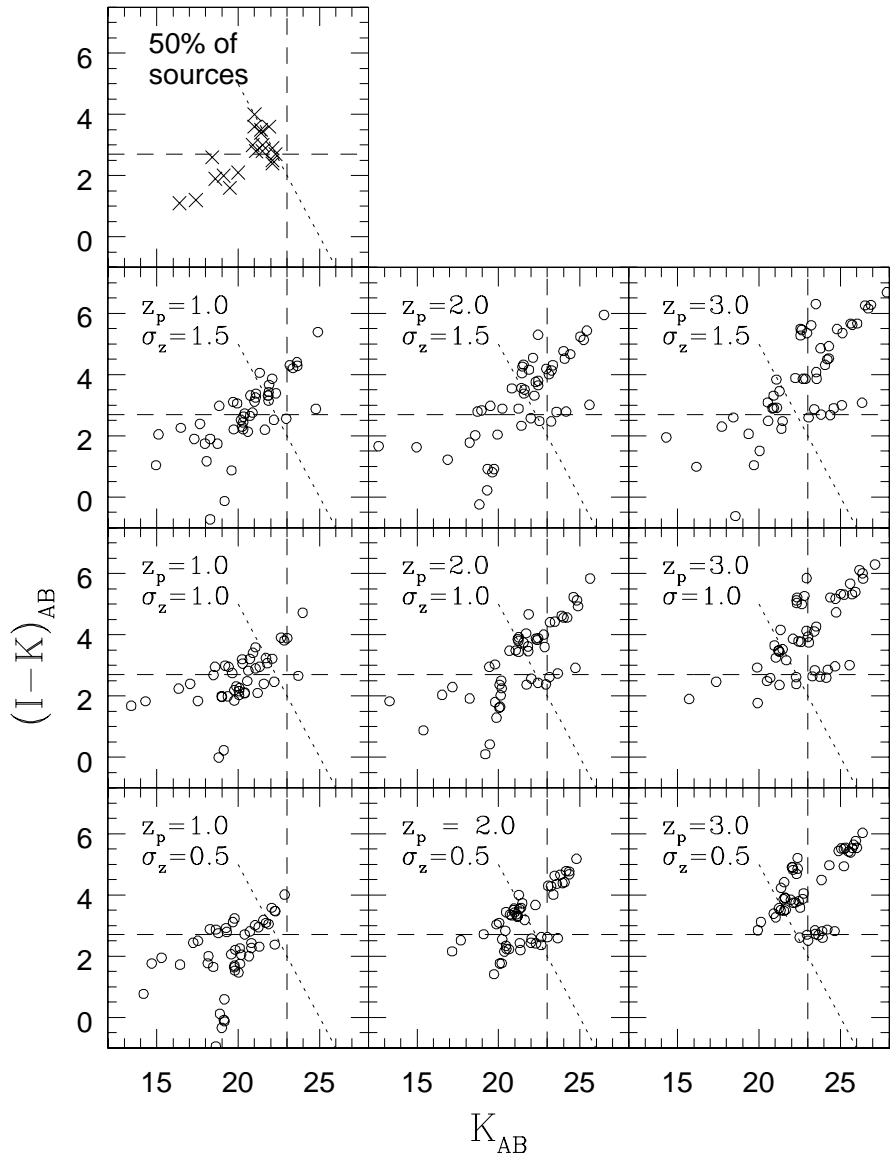}
\figcaption{ Sample 
colour-magnitude diagrams from the simple Gaussian redshift models. Each panel is labeled with 
a mean redshift and standard deviation of the redshift distribution.  The open circles correspond to a 
artificial SCUBA data set of 50 sources.  The sources have been randomly chosen from the four basic 
templates (14.1, 14.13, 14.18, and 3.10) and  placed in redshift space, according the the 
model redshift distribution.  Also shown, (top panel) are the plausible SCUBA source identifications
 discussed in \S5.3.  The dashed lines show our approximate $K$-band limit and the $(I-K)$ colour corresponding to an ERO. The dotted line corresponds to the limit at which an object is detected in $K$ but not in $I$.   These 
plots are meant to show the general neighborhood inhabited by each model.  As the 
sample is only 50 there will be some stochastic scatter in the plots.  \label{model1}}
\end{figure}

Figure \ref{model1} shows sample NIR colour-magnitude diagrams for a variety of models.  For these plots, 
artificial catalogues of 50 SCUBA sources 
have been produced (the same number as our  catalogue).

From these plots a few basic conclusions may be drawn.   It is immediately obvious that all three redshift distributions centered on $z$ = 1 have a surplus of identifications at bright $K$-magnitudes, compared to the real data. The implication that the median redshift must lie at $z> $ 1 has been  well established in the previous section as well as in the work of many previous papers, including our own (see previous section and references therein).
While a low-redshift distribution would produce too many bright identificaitons,  a narrow redshift distribution centered at $z$ = 3 has the opposite problem as all the counterparts would lie beyond our $K$-band limit, and  yet we claim that for $\sim$50\% of our sources we  have detected  $K$-band counterparts.  Thus, a  broad  redshift distribution (in this case 
$\sigma_z$ = 1.0 - 1.5 ) centered at $z\sim$ 2-3 is most consistent 
with these identifications.\

\citet{ivi00b} have put forth  a phenomenological classification scheme for the SCUBA sources 
based on their optical/NIR magnitudes.  Class-0 objects are very faint in both $I$, ($I_{vega}\ga$ 26), and 
$K$, ($K_{vega}\ga$ 21) and so their optical/NIR counterparts are below the detection limit of typical
observations.  Class-I objects have $K_{vega}\la$ 21 and $I_{vega}\ga$ 26 and are therefore classified as EROs, 
detected 
only at 
$K$.  Class-II objects are relatively bright in both bands with $I_{vega}\la$ 26 and $K_{vega}\la$ 21.  A fourth class 
is also possible in which an object is detected at $I$ but not at $K$.  Though SCUBA sources are 
expected to be red this scenario would be possible if some 
had SEDs similar to IRAS F12112+0305, the relatively blue SED from \citet{tre99} (see Figure 
\ref{colmag}).   This broad classification scheme roughly divides the population into different phenomenological catalgories.  Class-II systems likely lie within the low-redshift end of the population and may have been forming stars long enough to have built up significant stellar populations.  Class-I sources are all classified as EROs. Class-O sources likely represent the most extreme systems, either at very high-redshift or with immense amounts of dust extinction.   Laid out in this way, this classification scheme may represent an age sequence \citep{sma02}.

 Though it would be preferable to apply this classification scheme directly to our identifications, the different $I$ and $K$ follow-up depths in these data  and those of \citet{ivi00b} make this impossible.
Therfore we introduce a  second classification system based on the specific flux limits of this 
work in order to 
directly compare the results of the models with  the catalogue. We denote three Classes which are 
analogous to 
Ivison's Classes and use the following nomenclature: \\

{\footnotesize
\noindent ClassW-O:  $I_{AB}\ga$ 25  ($I\ga$ 25.45); $K_{AB}\ga$  23 ($K\ga$ 21.2) \

\noindent ClassW-I :  $I_{AB}\ga$ 25 ($I\ga$ 25.45); $K_{AB}\la$  23 ($K\la$ 21.2) \

\noindent ClassW-II:  $I_{AB}\la$ 25 ($I\la$ 25.45); $K_{AB}\la$  23 ($K\la$  21.2) } \\

\noindent In the remaining discussion we refer to Ivison's classes as ClassI and our own as ClassW. 

We have generated 5000 artificial SCUBA sources and these have been grouped,  for each model realization, into their 
respective Class. These results are presented in Tables 6.3 and 6.4.

The number of objects in ClassW-0 obviously increases with increasing $z_p$, since  as 
the objects are shifted to a higher median redshift a larger fraction drop below the detection limit.  The 
number 
of Class-II objects  correspondingly decreases.  At 
$z_p$ = 1,  70-90\% of the  NIR counterparts are classified as ClassW-II and would be detected 
in our data (and typical optical or NIR surveys), at $z_p$ = 2 approximately 
50\%-60\% (depending on the width of the distribution), and at $z_p$ = 3, 10\% - 30\%.  In 
most  cases the number of objects classified at ClassW-I is small.  This is a consequence of both 
the observational limits at $I$ and $K$ and the location of the redshift tracks of the four template 
galaxies.  
Once a galaxy becomes undetectable at $I$ the redshift window over which it is visible in $K$ is 
comparatively short.  The exception to this is the highest redshift set of simulations, $z_p$ = 3, in 
particular the narrow redshift distribution.  In this scenario most galaxies are located in  the 
neighborhood of 
$z\sim$ 3 where  these template galaxies are expected to be visible at $K$ but not at $I$ (see Figure 
\ref{tracks}). Thus, the  
template galaxies tend to ``bunch up'' within  ClassW-I.  

We may set some  strict limits on the fraction of real CUDSS SCUBA sources in each group. There are 11 radio-detected sources with counterparts detected in both $I$ and $K$ and therefore at least 22\% of the sample are ClassW-II objects.   Including the likely identifications discussed in \S 5.3 increases this fraction to 32\%. There is one radio source  that is detected with $K_{AB}\leq$ 23 but not in $I$, plus three likely identifications. This sets a lower limit on the fraction of sources that are ClassW-I at 2-8\%.    There are five empty fields in the 3\h field, and two radio sources with $I_{AB}\geq$ 25 and $K_{AB}\geq$ 23.  This sets a lower-limit on the fraction of ClassW-O sources at 14\%.  We can also set an upper-limit on the number of objects in this group from the radio detections at $<$ 76\% and from all the likely identifications, $<$ 60\%.   Comparing these limits with Table 7 we see that a high median redshift of $z\sim$2-3, with a broad redshift distribution produces the best reproduction of the data.

 These fractions are all  broadly consistent with the classification of sources  in two other systematically selected samples, the ``8 mJy Survey'' \citep{ivi02} and the deeper  lensing survey  of \citet{sma02}.  If the separation of objects in ClassII-Class0 is the result  of an evolutionary sequence, and surveys of different flux-limits are selecting different redshift regimes \citep{cha02,ivi02} one might expect to see differences between the fraction of objects in each Class for different surveys.  Though there is no significant difference in the number of ClassII objects we have in our catalogue, compared to other groups, we do have an unusually large number of very bright, low-redshift, counterparts.  This would be expected if we are biased towards a population at lower redshift than shallower surveys.  Indeed, none of our low-redshift objects would have been detected in a survey with an 8 mJy flux-limit. Moreover, though observations  which reach deeper flux-limits than the CUDSS would also be expected to be sensitive to these lower-redshift sytems, they survey much smaller sky areas \citep{hug98} and the bulk are lensing surveys \citep{sma97,cow02},  designed to pick up lensed SCUBA sources behind galaxy clusters. Thus, perhaps they simply have not have covered enough of the low-redshift universe to detect these objects, which make up a small fraction of the overall SCUBA population.

\subsection{SCUBA sources and optical/NIR-selected poplations }

Of key interest  is the overlap between the SCUBA sources and the ERO population. Given only our secure radio identifications, at least 10\% of the CUDSS sources are associated with bright EROs (that is, ClassW-II EROs).  Including the likely identifications raises this fraction to 22\%. This sets a strict upper-limit to the number of bright EROs since there are no EROs within 8\arcsec of a SCUBA source that were not chosen as the best (and likely)  identification. This agrees very well with the estimate of \citet{ivi02} that approximately one third of the ``8 mJy Survey'' sources are classified as EROs. 

 However, in Table 7 we see the fraction of ClassW-II EROs is not very sensitive to the redshift distribution. Of more interest, but harder  to estimate, is the total number of SCUBA sources associated with EROs, over all magnitudes. Between  $z_{peak}\sim$ 2 and $z_{peak}\sim$ 3 there is a substantial difference in the expected number of total  ERO counterparts, ranging from 60\% to 90\%.   Unfortunately, bulk of these  lie at faint $K$ magnitues where the  ERO population has been sparseley studied.  We may place an upper-limit on the true fraction based on the 8 radio sources which are not EROs ($\leq$ 84\%) and the lower-limit of 22\% from the number of bright ERO identifications. 

The surface density of EROs is not well constrained, due to small area coverage and the strong clustering of the population.  However, if we take $\sim$ 2-3 arcmin$^{-2}$ as a reasonable surface density for $K_{AB}\geq$ 23 EROs \citep{smi02,weh02}, and note that SCUBA galaxies with $S_{850{\mu}m} \geq$ 3 mJy have a surface density of $\sim$ 0.5 arcmin$^{-2}$ then as many as 10\% of EROs to this  $K$-limit may be submillimeter bright.

\begin{figure}
\plotone{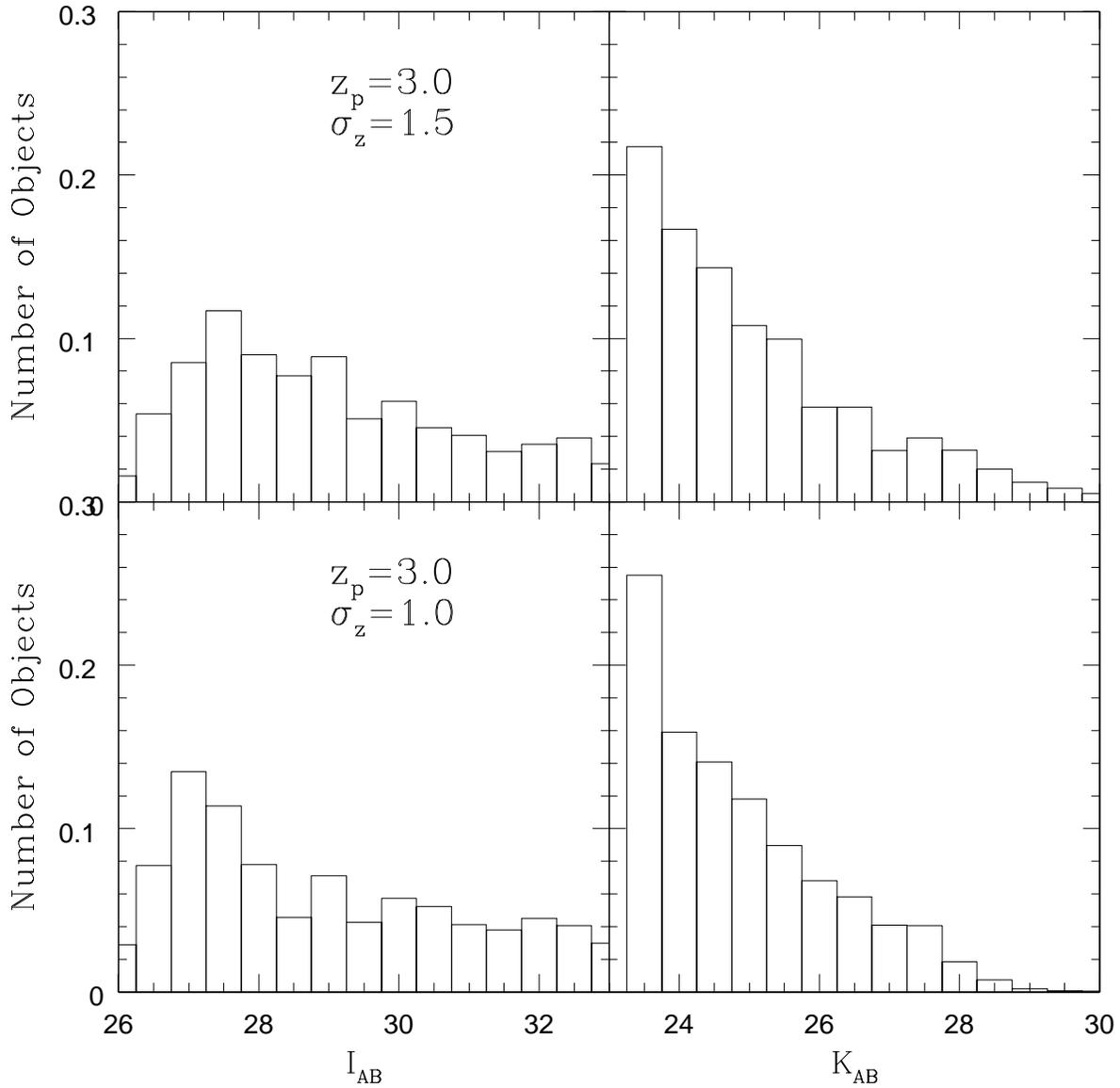}
\figcaption{This plot 
shows the distribution $I$ and $K$ magnitudes of sources classified as Group 0 
for two high-redshift distributions (normalized to the number of objects in the group). \label{group0}}
\end{figure}

During the  discussion in the previous sections we concluded that the  remaining $\sim$ 50\% of the CUDSS sources which are unidentified do not 
have 
identifications present in our $K$-image  but all lie in 
ClassW-0, with $K_{AB}\ga$ 23 and $I_{AB}\ga$ 26. 
In Figure \ref{group0}  we show the distribution of $I$ and $K$ magnitudes for $z_{peak}$=3 and 
$\sigma_z$ = 
1.0, 1.5.  It is immediately obvious that $K$ observations remain the method of choice over $I$ for detecting 
these 
counterparts.  Roughly 80\% of the counterparts have $K_{AB}\leq$ 26 ($K_{vega}\leq$24.2) while at $I$ the 
objects 
are very evenly distributed over a wide range in magnitude.

 \citet{sma02} have reported a median magnitude of $I \ge$ 26 for the counterparts in 
their  lensed sample, clearly implying the 
bright submillimeter sources are distinct from optically selected  populations.   The 
optical  counterparts  in our survey are also faint, though they  have a slightly brighter median magnitude than found by 
\citet{sma02}.    Roughly 40\% of our sample have secure or likely identifications with $I_{AB} \leq$ 25 ($I_{vega} \leq$ 24.5). Therefore  our median $I$-magnitude will surely be  $I_{AB}\la$ 26.
The ``8 mJy survey'' has also reported an abundance of relatively bright optical and NIR counterparts.  
With deep 1.4 GHz imaging \citep{ivi02} they have secured positions for 18/30  sources in their 
catalogue.  Roughly 90\% of these have optical counterparts with $I<$ 25 ($I_{AB}<$25.5),  which 
implies $\sim$50\% of their entire sample  is brighter than this limit.  

The Smail et al. lensing project and the ``8 mJy survey''  target slightly different 850\micron \ flux 
regimes and therefore one might expect a variation in optical/NIR  magnitudes  between the two samples.
However,   different behaviours of the $K$-corrections at 850\micron \ and in the optical/NIR, means the relationship between optical or NIR magnitude and submillimeter flux is a combination of both the intrinsic properties of the galaxies and the redshift distribution.If  intrinsic (and 
hence observed) submillimeter flux  increases with redshift \citep{cha02,ivi02}, the  optical/NIR properties 
were 
uniform across the submillimeter population,  this would lead predominantly fainter counterparts in these 
bands for the brightest SCUBA sources.    

\begin{figure}
\plotone{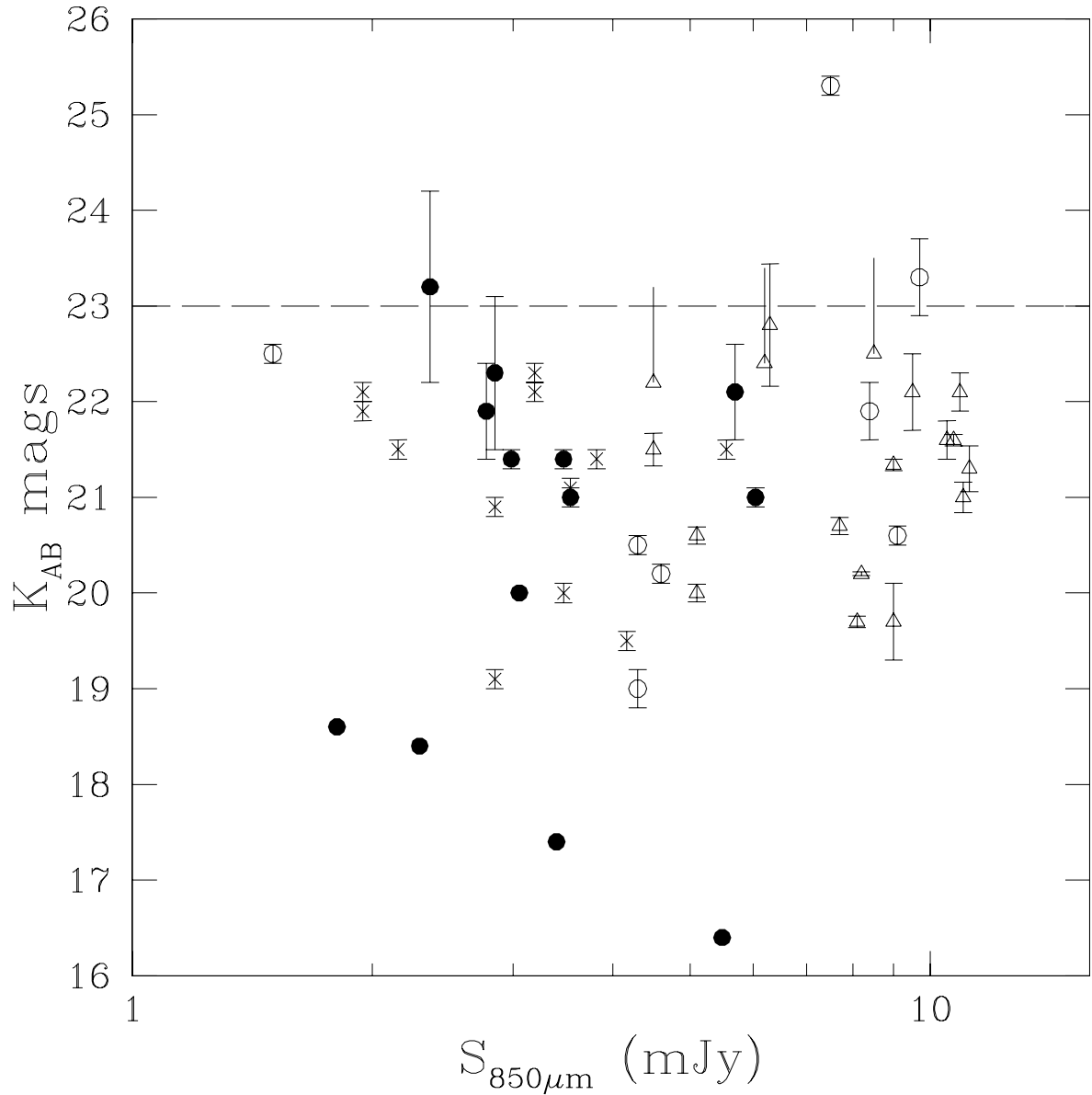}
\caption{ This plot 
shows 
the $K$-magnitude versus 850\micron \ flux for the best identifications in Table 8.  
The solid circles represent the radio-identified sources of the both fields of the CUDSS and the crosses correspond to likely identifications from \S 5.3.  The open circles represent the 
identifications from \citet{sma02} and the open triangles the radio identifications of \citet{ivi02}. \label{k850}}
\end{figure}

\citet{sma02}  found  that  though the submillimeter sources  possess a wide range of optical/NIR 
properties, faint SCUBA sources ($S_{850\micron} < $ 4 mJy)  are typically identified with very faint 
NIR counterparts (albeit dealing with small numbers of sources).  In Figure \ref{k850} we plot the $K$-magnitude of the identifications  in  
this work against 
the 850\micron \ flux densities, as well as the results of Smail et al.  and \citet{ivi02}.   The 13 radio sources with $K$-band detections are represented by the solid circles, the Smail et al sources are denoted by the open 
circles and the Ivsion et al. by the open triangle.   It is immediately seen that some fainter SCUBA sources in this sample have very bright NIR 
counterparts.   Also plotted (crosses) are the likely identifications from \S 5.3 which also show  a broad 
range of 
$K$-magnitudes (a factor of 25 in flux) for a very small range of submillimeter flux (a factor of 3 in 
flux). Thus, we see no evidence in these data  for a trend in $K$-magnitude with submillimeter flux, and in particular our faint submillimeter sources are not restricted to faint $K$-band counterparts.  However, since this does not appear to be the case for the other two surveys this again highlights the possibility that different surveys are sensitive to different objects or redshift regimes.

\section{Conclusions}

(i) Using multi-wavelength data, but focusing on the NIR, we have selected identifications for submillimeter 
sources in the CUDSS 14\h field. We argue that 
our NIR algorithm is just as effective at correctly identifying the counter-parts as radio follow-up, with 
minimal contamination from incorrect 
identifications.  

(ii) Using the securely identified radio sources as templates we select other likely identifications, based on 
common properties. We discuss these 
identifications in detail and find  that many are very red and show unusual morphologies.  

(iii) We claim that we have identified $\sim$50\% of the submm sources in the CUDSS sample.  Almost all  of these 
identifications have $I_{AB} <$ 25, and 
therefore many are within reach of deep spectroscopic observations on 8-m class telescopes. 

(iv) We present the spectroscopic and estimated redshifts for the identifications.  We use the 
1.4 GHz-to-850\micron \ flux ratio, and the \citet{yun02} 
template SED fitting technique as our main methods.  We also use the radio sources as templates to 
estimate NIR photometric redshifts for the 
non-radio detected identifications.

(v) We place a lower-limit on the median redshift  of this population of $z \geq$ 1.4.  We argue that 
4\%-10\% of the sources lie at $z <$ 1.0, and at least 40\% lie at $z <$ 3.

(vi)  Using simple models of the redshift distribution we argue that these data are consistent with a high 
median redshift ($z\sim$ 3.0) and a broad 
distribution (${\sigma}_z$ = 1.0 - 1.5 for a Gaussian distribution).  

(vii) To a NIR limit of $K_{AB} <$ 23.0,  $\sim$ 10-22\%  of our identifications are classified as EROs. 
However, we argue that EROs truly make up 
$\sim$60-84\% of the  sample, with the bulk of these found below our $K$-band limit. Most of the remaining 
identifications, though not technically 
classified as EROs will still be very red. Hence deep $K$-band imaging will continue to be an important 
tool in the follow-up of these sources.  

(viii) Our  submillimeter sample, which spans a range of 3 $\la S_{850_{\mu}m} \la $ 8 mJy, shows no 
correlation between $K$ magnitude and submillimeter flux.

\

{\it Acknowledgments}

We are grateful to the many members of the staff of the Joint Astronomy Centre who have helped us with
this project.  Research by Tracy
Webb was supported by the National Sciences and Engineering Council of Canada, the Canadian National
Research Council,  the Ontario Graduate
Scholarship Program, the Walter C. Sumner Foundation, and the NOVA Postdoctoral Fellowship program.       
At the start of this program, research by Simon Lilly was
supported by the National Sciences and Engineering Council of Canada and by the
Canadian Institute of
Advanced Research. Research by
Stephen Eales, David Clements, Loretta Dunne and Walter Gear is supported by the Particle Physics and
Astronomy Research Council.  Stephen Eales
also acknowledges support from Leverhulme Trust.  The JCMT
is operated by the Joint Astronomy Centre on behalf of the UK Particle Physics and Astronomy Research
Council, the Netherlands Organization
for Scientific Research and the Canadian National Research Council.   \

\clearpage

\onecolumn
\begin{deluxetable}{lllll}
\tabletypesize{\scriptsize}
\tablecaption{The 14-Hour Field Catalogue}
\tablewidth{0pt}
\tablehead{
\colhead{Name} & \colhead{Old Name} & \colhead{R.A. and Decl. (J2000.0) } & \colhead{S/N} & 
\colhead{$S_{850{\mu}m}$ mJy}}

\startdata
CUDSS 14.1 & 14A & 14 17 40.25, 52 29 06.50  & 10.1 & 8.7 $\pm$ 1.0 \\
CUDSS 14.2 & 14B & 14 17 51.70, 52 30 30.50 & 6.3 & 5.5 $\pm$ 0.9 \\
CUDSS 14.3 &  ...   & 14 18 00.50, 52 28 23.50 & 5.4 & 5.0 $\pm$ 1.0 \\
CUDSS 14.4 & ... & 14 17 43.35, 52 28 14.50 & 5.3 & 4.9 $\pm$ 0.9 \\
CUDSS 14.5 & ... & 14 18 07.65, 52 28 21.00 & 4.5 & 4.6 $\pm$ 1.0 \\
CUDSS 14.6 &  ... & 14 17 56.60, 52 29 07.00 & 4.2 & 4.1 $\pm$ 1.0 \\
CUDSS 14.7 & ... & 14 18 01.10, 52 29 49.00 & 3.2 & 3.2 $\pm$ 0.9 \\
CUDSS 14.8 & 14E & 14 18 02.70, 52 30 15.00 & 4.0 & 3.4 $\pm$ 0.9 \\
CUDSS 14.9 & ... & 14 18 09.00, 52 28 04.00 & 4.1 & 4.3 $\pm$ 1.0 \\
CUDSS 14.10 & ... & 14 18 03.90, 52 29 38.50 & 3.5 & 3.0  $\pm$ 0.8  \\
CUDSS 14.11 & ...  & 14 17 47.10, 52 32 38.00 & 3.5 &  4.5 $\pm$ 1.3 \\
CUDSS 14.12 & ... & 14 18 05.30, 52 28 55.50 & 3.4 & 3.4 $\pm$ 1.0 \\
CUDSS 14.13 & ... & 14 17 41.20, 52 28 25.00 & 3.4 & 3.3 $\pm$ 1.0 \\
CUDSS 14.14 & ... & 14 18 08.65, 52 31 03.50 & 3.3 & 4.6 $\pm$ 1.3 \\
CUDSS 14.15 & ... & 14 17 29.30, 52 28 19.00 & 3.1 & 4.8 $\pm$ 1.5 \\
CUDSS 14.16 & ... & 14 18 12.25, 52 29 20.00 & 3.7 & 4.7 $\pm$ 1.4 \\
CUDSS 14.17 & ... & 14 17 25.45, 52 30 44.00 & 3.3 & 6.0 $\pm$ 2.1 \\
CUDSS 14.18 & 14F  & 14 17 42.25, 52 30 26.50 & 3.0 & 2.6 $\pm$ 0.9 \\
CUDSS 14.19 & ... & 14 18 11.50, 52 30 04.00 & 3.0 & 3.9 $\pm$ 1.3 \\
CUDSS 14.20 & ...  & 14 17 50.40, 52 31 04.00 & 3.0 & 2.8 $\pm$ 0.9  \\
CUDSS 14.21 & 14D & 14 18 02.30, 52 30 51.50 & 3.0 & 2.8 $\pm$ 0.9 \\
CUDSS 14.22 & ... & 14 17 55.80, 52 32 04.00 & 3.0 & 2.9 $\pm$ 1.0 \\
CUDSS 14.23 & ...  & 14 17 46.30, 52 33 24.00 & 3.0 & 2.8 $\pm$ 0.9 \\

\enddata
\end{deluxetable}

\begin{deluxetable}{lclrcl}
\tabletypesize{\scriptsize}
\tablecaption{SCUBA Sources with 1.4 GHz Detections \label{radiotab} }
\tablewidth{0pt}
\tablehead{
\colhead{SCUBA name} & \colhead{Radio ID} & \colhead{R.A. and Dec. J2000} & \colhead{$S_{1.4GHz}$($\mu$Jy)} & \colhead{Offset (\arcsec)} & \colhead{ 
$P^\prime$} 
}

\startdata
14.1 & 15V18 & 14 17 40.32, 52 29 05.9 & 134 $\pm$ 20 & 0.9 & 6.0 $\times$ 10$^{-4}$  \\
14.3 & 15V23 & 14 18 00.50, 52 28 20.8 & 138 $\pm$ 21 & 2.7 & 8.6 $\times$ 10$^{-3}$  \\
14.9  & 15V67 & 14 18 09.30, 52 28 03.0 & 67 $\pm$ 17 & 1.0 & 1.1 $\times$ 10$^{-3}$  \\
14.13 & 15V23 & 14 17 41.81, 52 28 23.4 & 109 $\pm$ 18 & 5.8 & 2.9 $\times$ 10$^{-3}$ \\
14.18 & 15V24 & 14 17 42.08, 52 30 25.2 &  235 $\pm$ 35 & 2.0 & 2.4 $\times$ 10$^{-4}$  \\
14.19\tablenotemark{a} & ... & 14 18 11.02, 52 30 11.60 & 82 $\pm$ 19 & 8.5 & 2.5 $\times$ 10$^{-2}$ \\
\enddata
\tablenotetext{a}{This identification is beyond our nominal search radius.  Though we expect 5-10\% of the identifications to  have offsets  of $>$ 8{\arcsec} we cannot regard this identification as secure}
\end{deluxetable}

\begin{deluxetable}{lclcl}
\tabletypesize{\scriptsize}
\tablecaption{SCUBA Sources with ISO 7\micron \ and 15\micron \ Detections \label{isotab} }
\tablewidth{0pt}
\tablehead{
\colhead{SCUBA name} & \colhead{ISO Number} & \colhead{R.A. and Dec. J2000}  &\colhead{Offset (\arcsec)} & \colhead{ 
$P^\prime$} 
}

\startdata
14.13 & 0  & 14 17 41.81, 52 28 23.0 & 5.9 & 0.03 \\
14.17 & 195 & 14 17 24.36, 52 30 46.5 & 10.3 & 0.08  \\
14.18  & 5 & 14 17 42.04, 52 30 25.7 & 2.1 & 0.0036  \\
\enddata
\end{deluxetable}

\begin{deluxetable}{lllllcl}
\tabletypesize{\scriptsize}
\tablecaption{Optical and Near-Infrared Identifications \label{idtab}}
\tablewidth{0pt}
\tablehead{
\colhead{Name} & \colhead{CFRS name} & \colhead{$K_{AB}$} & \colhead{$(I-K)_{AB}$} & \colhead{R.A. and Dec. J2000} & \colhead{Offset (\arcsec)}  & \colhead{$P^\prime$} 
}

\startdata
14.1\tablenotemark{a} & ... & 21.0 $\pm$ 0.1 & 3.6 $\pm$ 0.4 & 14 17 40.32, 52 29 05.90  & 0.9 & 0.041 \\
14.2 & ... & 21.4 $\pm$ 0.1 & 3.5 $\pm$ 0.4 & 14 17 51.43, 52 30 25.40 & 5.6 &  0.15  \\
14.3\tablenotemark{a} & ... & 21.4 $\pm$ 0.1 & 3.4 $\pm$ 0.4 & 14 18 00.40, 52 28 20.30  & 3.3  & 0.076   \\
14.4 & ... & 21.5  $\pm$ 0.1& 1.8 $\pm$ 0.3  & 14 17 43.63, 52 28 18.90  & 5.1 &  0.32 \\
14 .5 & ... & 22.3 $\pm$ 0.1 & 2.7 $\pm$ 0.4 & 14 18 08.19, 52 28 21.00 & 4.9 & 0.28  \\
14.6\tablenotemark{b}& ...   & 24.0  $\pm$ 0.2 & 0.2 $\pm$ 0.5 & 14 17 56.21, 52 29 01.50 & 6.5 & 0.79 \\
14.7\tablenotemark{b} & ... & 22.2 $\pm$ 0.1 & 2.8 $\pm$ 1.0 & 14 18 00.87, 52 29 49.20 & 2.1  & 0.11 \\
14.8\tablenotemark{b} & ... & 22.3  $\pm$ 0.1 & 2.7 $\pm$ 1.0  & 14 18 02.87, 52 30 11.11 & 4.2  & 0.25 \\
14.9\tablenotemark{a} & ... &  21.0 $\pm$ 0.1 & $\geq$ 4.0 & 14 18 09.00, 52 28 03.80 & 3.3  & 0.034 \\
14.10\tablenotemark{b} & ... & 22.1 $\pm$ 0.1  & 1.0 $\pm$ 0.2 & 14 18 03.97, 52 29 34.15 & 4.4  & 0.42 \\
14.11 & 03.0986 & 17.3  $\pm$ 0.1& 0.0 $\pm$ 0.1  & 14 17 47.26, 52 32 42.17&  4.4  & 0.078 \\
14.12 & ... & 22.6  $\pm$ 0.1& 1.3 $\pm$ 0.5 & 14 18 05.17, 52 28 50.40 & 5.2   & 0.55 \\
14.13 & 03.1157\tablenotemark{a} & 18.4 $\pm$ 0.1 & 2.5 $\pm$ 0.4 & 14 17 41.81, 52 28 22.99&  5.8  & 0.010 \\
14.14 & ... & 22.1 $\pm$ 0.1 & 2.5 $\pm$ 0.4 & 14 18 09.45, 52 31 05.25& 7.4  & 0.41 \\
14.15 & ...  & 20.7  $\pm$ 0.1 & 1.5 $\pm$ 0.1  & 14 17 29.89,  52 28 21.20 & 5.7  & 0.45 \\
14.16 & 03.0310 & 20.7 $\pm$ 0.1  & 0.4 $\pm$ 0.1 & 14 18 11.93,  52 29 14.76&  6.0  & 0.37 \\
14.17 & ... & 19.5 $\pm$ 0.1 & 1.6 $\pm$ 0.2  & 14 17 24.90, 52 30 42.00&  5.3 &  0.22 \\
14.18 & 03.1139\tablenotemark{a} & 18.6 $\pm$ 0.1 & 2.4 $\pm$ 0.3 & 14 17 42.04, 52 30 25.69 & 2.1  & 0.001 \\
14.19 & ... & 22.8  $\pm$ 0.1 & 2.0 $\pm$ 0.9 & 14 18 11.22, 52 30 02.30&  3.0  & 0.30 \\
14.20 & ... & 23.2  $\pm$ 0.1 & 2.1 $\pm$ 0.8 & 14 17 50.50, 52 31 01.00&  3.1  & 0.31 \\
14.21 & ... & 23.2 $\pm$ 0.1 & 0.4 $\pm$ 0.3 & 14 18 02.24, 52 30 48.20& 3.3  & 0.47 \\
14.22 & ... & 22.1 $\pm$ 0.1 & 2.4 $\pm$ 0.8  &  14 17 54.91, 52 32 08.24 & 6.9  & 0.49 \\
14.23 & ... & 21.9 $\pm$ 0.1 & 3.6 $\pm$ 1.0 & 14 17 46.10, 52 33 22.20& 2.5 & 0.060 \\
\enddata
\tablenotetext{a}{This object has been detected at 5 GHz}
\tablenotetext{b}{This object also has an LBG association. Please see Table \label{lbgtab}.}
\end{deluxetable}

\begin{deluxetable}{lllll}
\tabletypesize{\scriptsize}
\tablecaption{Lyman-Break Galaxy Associations \label{lbgid}}
\tablewidth{0pt}
\tablehead{
\colhead{SCUBA name} & \colhead{LBG Identification} &\colhead{Offset (arcsecs)} & \colhead{ 
$P^\prime$} & \colhead{Note}
}

\startdata
14.6 & CFDF 64601 & 6.2 & 0.019 & Same ID as Table \ref{idtab}. \\
14.7 &Steidel  West2 MMD13 & 7.8 & 0.096 & Different ID than Table \ref{idtab}. Confirmed QSO. \\
14.8  & Steidel MMD75 & 7.2 & 0.081 & Different ID than Table \ref{idtab}. \\
14.10 & Steidel MMD63 & 6.7 & 0.072 & Different ID than Table \ref{idtab}. \\
\enddata
\end{deluxetable}

\begin{deluxetable}{llll|llll}
\tabletypesize{\scriptsize}
\tablecaption{The Redshift Estimates \label{redshifts}}
\tablewidth{0pt}
\tablehead{
\colhead{SCUBA} & \colhead{Millimetric} & \colhead{Spect.}  & \colhead{NIR}  &  \colhead{SCUBA} & 
\colhead{Millimetric} & \colhead{Spect.}  & 
\colhead{NIR}  \\
\colhead{Name}  & \colhead{Redshift} & \colhead{Redshift} & \colhead{Redshift\tablenotemark{a}} & 
\colhead{Name}  & 
\colhead{Redshift} & \colhead{Redshift} & 
\colhead{Redshift\tablenotemark{a}}  

}

\startdata

 3.1 & $>$ 2.3  &  -- & --  &  3.26 & $>$ 1.4  & -- & -- \\
 3.2 & $>$ 1.6  & --  &  -- &  3.27 &  1.3 $\pm$ 0.3  & --  & -- \\ 
 3.3 &$>$ 1.9  & -- & --   &  14.1 & 2.4 $\pm$ 0.5  & -- & 2.9 $\pm$ 0.5   \\
 3.4 & $>$ 2.0  & -- & 0.9 $\pm$ 0.2 &  14.2 & $>$ 1.4 &  -- & 2.9 $\pm$ 0.5  \\
 3.5 & $>$ 1.5  & -- & -- &  14.3 & 2.0 $\pm$0.4 &  --  & 2.6 $\pm$ 0.6  \\
 3.6 & 1.3 $\pm$ 0.3   & -- & --  &  14.4 & $>$ 1.5 &  -- & -- \\
 3.7 & 2.0 $\pm$ 0.4  & --  & $>$ 1.0 &  14.5  & $>$ 1.5 &  --  & 3.1 $\pm$ 0.8  \\
 3.8 & 0.29 $\pm$ 0.08  & 0.0880  & -- &   14.6 & $>$ 1.4 &  -- & -- \\
 3.9 &   $>$ 1.7  & -- & --  &   14.7 & $>$ 1.2 & -- & --\\
 3.10 & 0.8 $\pm$ 0.2 & 0.176  & -- &  14.8 & $>$ 1.1 &  -- & -- \\
 3.11 & $>$ 1.6  & -- & --  &  14.9 & 1.7 $\pm$0.6   &  -- & -- \\
 3.12 & $>$ 1.6   & -- & -- &  14.10 & $>$1.2 &  -- & --  \\
 3.13 & $>$ 1.5   &  -- & 0.8 $\pm$ 0.5 &  14.11 &  $>$ 1.1    & -- & -- \\
 3.14 & $>$ 1.6   & -- & -- &   14.12 & $>$ 1.3 &  -- & -- \\
 3.15 & 0.58 $\pm$ 0.2 &  -- & -- &  14.13 &  1.4 $\pm$  0.5 & 1.15    & -- \\
 3.16 & $>$ 1.2  &   -- &  -- &  14.14 & $>$ 1.22 &  -- & 1.9 $\pm$ 0.9  \\
 3.17 & 1.1 $\pm$ 0.3   & -- & 1.1 $\pm$ 0.4 &  14.15 & $>$ 1.2 &  -- & -- \\
 3.18 & $>$ 1.5  & --& -- &   14.16 & $>$ 1.4 &  -- & -- \\
 3.19 & $>$ 1.4   & -- & -- &   14.17 & $>$ 1.6 &  -- & --\\
 3.20 & $>$ 1.4  & -- & -- &   14.18 & 1.4 $\pm$ 0.3  & 0.660   & -- \\
 3.21 & $>$ 1.5  & -- & --&   14.19 & $>$1.2 &  -- & -- \\
 3.22 & $>$ 1.3   & -- & 1.0 $\pm$ 0.2  &  14.20 & $>$ 1.3 &   -- & -- \\
 3.23 & $>$ 1.7  & --  & -- &   14.21 & $>$ 1.3 &  -- & -- \\
 3.24 & 1.0 $\pm$ 0.3  &  -- & 0.9 $\pm$ 0.2  &  14.22 & $>$ 1.3 &  -- & 1.8 $\pm$ 0.6  \\
 3.25 &  0.3 $\pm$ 0.09  & -- & --   &   14.23 & $>$ 1.3  & -- & 3.1 $\pm$ 0.6 \\

\enddata
\tablenotetext{a}{Using the optical/NIR templates from \S 5.}
\end{deluxetable}

\begin{deluxetable}{llrrrrr}
\tabletypesize{\scriptsize}
\tablecaption{Model Redshift Distributions (Using Limits of this Work)}
\tablewidth{0pt}
\tablehead{
\colhead{$z_{peak}$} & \colhead{${\sigma}_z$} & \colhead{ClassW-0} & \colhead{ClassW-I} & \colhead{Class-II} & \colhead{ClassW-II EROs} & 
\colhead{Total Eros}
}

\startdata
1.0 & 0.5 & 3 \% & 8 \% & 89 \% & 18 \% & 29 \% \\
1.0 & 1.0 & 8 \% & 11 \% & 81 \% & 21 \% & 38 \% \\ 
1.0 & 1.5  & 16 \% & 13 \% & 71 \% & 20 \% & 46 \% \\
\hline
2.0 & 0.5 & 24 \% & 16 \% & 60 \% & 34 \% & 68 \% \\
2.0 & 1.0 & 26 \% & 21 \% & 53 \% & 23 \% & 65 \% \\
2.0 & 1.5 & 31 \% & 20 \% & 49 \% & 18 \% & 66 \% \\
\hline
3.0 & 0.5 & 47 \% & 41 \% & 12 \% & 10 \% & 94 \% \\ 
3.0 & 1.0 & 48 \% & 29 \% & 23 \% & 14 \% & 87 \% \\
3.0 & 1.5 & 51 \% & 22 \% & 27 \% & 13 \% & 82 \% \\
\enddata
\end{deluxetable}

\begin{deluxetable}{llrrrrr}
\tabletypesize{\scriptsize}
\tablecaption{Model Redshift Distributions (Using Classification of Ivison)}
\tablewidth{0pt}
\tablehead{

\colhead{$z_{peak}$} & \colhead{${\sigma}_z$} & \colhead{Class 0} & \colhead{Class I} & \colhead{Class II} 
& \colhead{Class-II EROs} & 
\colhead{Total Eros}
}

\startdata
1.0 & 0.5 & 4 \% & 1 \% & 95 \% & 24 \% & 29 \% \\
1.0 & 1.0 & 8 \% & 1 \% & 89 \% & 23 \% & 34 \% \\
1.0 & 1.5 & 13 \% & 2 \% & 82 \% & 21 \% & 37 \% \\ 
\hline
2.0 & 0.5 & 21 \% & 1 \% & 73 \% & 44 \% & 68 \% \\
2.0 & 1.0 & 23 \% & 4 \% & 69 \% & 34 \% & 64 \% \\
2.0 & 1.5 & 28 \% & 4 \% & 64 \% & 26 \% & 60 \% \\
\hline
3.0 & 0.5 & 39 \% & 11 \% & 38 \% & 36 \% & 94 \% \\ 
3.0 & 1.0 & 46 \% & 8 \% & 40 \% & 29 \% & 87 \% \\
3.0 & 1.5 & 48 \% & 6 \% & 42 \% & 24 \% & 81 \% \\

\enddata
\end{deluxetable}

\clearpage

\clearpage

\clearpage

\clearpage

\end{document}